\documentclass[twocolumn]{aastex701}

\usepackage{xcolor}
\usepackage{gensymb}

\usepackage{amsmath}
\usepackage{bm}

\begin{document}

\title{Coherent Radio Emission from Flux Rope Mergers in 3D Relativistic Reconnection}

\author[0000-0002-6013-4655]{Jasmine Parsons}
\affiliation{Department of Astrophysical Sciences, Princeton University, Princeton, NJ 08544, USA}
\email{jasmine.parsons@princeton.edu}

\author[0000-0002-1227-2754]{Lorenzo Sironi}
\affiliation{Department of Astronomy and Columbia Astrophysics Laboratory, Columbia University, New York, NY 10027, USA}
\affiliation{Center for Computational Astrophysics, Flatiron Institute, 162 5th Avenue, New York, NY 10010, USA}
\email{}

\author[0000-0001-9179-9054]{Anatoly Spitkovsky}
\affiliation{Department of Astrophysical Sciences, Princeton University, Princeton, NJ 08544, USA}
\email{}

\author[0000-0001-5121-1594]{Alexander Chernoglazov}
\affiliation{Department of Astrophysical Sciences, Princeton University, Princeton, NJ 08544, USA}
\affiliation{Institute for Advanced Study, Princeton, NJ 08540, USA}
\email{}



\begin{abstract}

Coherent emission from merging plasmoids in relativistic reconnection is the leading model for radio emission originating in the outer magnetosphere of pulsars such as the Crab, and a candidate emission mechanism for fast radio bursts. This process has only been studied in 2D, and whether it survives in 3D, where the dynamics of structures in the reconnection layer can differ substantially, is still an open question. Here we present the first study of coherent wave generation from flux rope mergers in 3D relativistic reconnection. We use particle-in-cell simulations of relativistic reconnection with outflow boundary conditions and guide fields $B_g/B_0 = \{0,0.1,0.3\}$, and measure the efficiency $\eta$ with which inflowing magnetic energy is converted into escaping coherent waves. We find $\eta \sim 7\times10^{-4}$ without a guide field, a factor of a few below 2D results, and $\eta\sim(1\text{--}2) \times10^{-3}$ at moderate guide fields, comparable to 2D with no guide field. We propose a physical picture for why the efficiency remains comparable to 2D, based on how reconnected field is transported away from the diffusion regions. The escaping waves are linearly polarized, with a polarization angle set by the direction of the upstream field. In the presence of a guide field, waves escaping from opposite sides of the reconnection layer therefore differ in polarization angle by $ \sim 2\arctan(B_g/B_0)$. Our results support relativistic magnetic reconnection as the origin of coherent radio emission from Crab-like pulsars, and strengthen its case as a potential emission mechanism for fast radio bursts.

\end{abstract}


\section{Introduction} \label{sec:intro}

Coherent radio emission is observed from a variety of astrophysical sources, most notably pulsars and magnetars \citep[][]{philippov2022review, kaspi2017}. Fast radio bursts (FRBs), extremely bright radio pulses lasting milliseconds, have in recent years emerged as a major new class of coherent emitters \citep[][]{lorimer2007,chime2019, petroff2019review}. Identifying the physical mechanisms behind coherent radio emission is a long-standing problem, because this emission arises from collective microscopic plasma behavior that is challenging to model analytically. The last few years have seen substantial progress in explaining coherent radio emission from pulsars \citep[e.g.,][]{philippov2019,philippov2020prl, cruz2021, bransgrove2023, chernoglazov2024}, enabled in large part by particle-in-cell (PIC) simulations, which capture the kinetic microphysics of the magnetospheric plasma and thus model the generation of coherent emission from first principles.

One such development concerns pulsar radio emission that originates in the outer magnetosphere rather than at the polar cap. In the Crab pulsar, as well as several other young energetic pulsars and many millisecond pulsars, the radio emission is phase-aligned with the high-energy emission \citep[][]{moffett1996, abdo2010,guillemot2012,johnson2014, eilek2016, kramer2026}. The high-energy emission is thought to originate in the current sheet beyond the light cylinder \citep[][]{cerutti2016, philippov2018}, suggesting that the radio emission is produced in the same region. The leading model for this radio emission invokes magnetic reconnection in the current sheet \citep[][]{lyubarsky2019, philippov2019}. In this model, the current sheet fragments through the tearing instability into a chain of plasmoids \citep[or, in 3D, flux ropes;][]{loureiro2007}. As plasmoids in the sheet merge with each other, they drive rapidly varying currents that launch fast magnetosonic (FMS) waves into the upstream plasma. These waves travel across the background field, and, as the plasma density declines away from the sheet, become freely propagating electromagnetic waves, i.e., coherent radio emission. This model naturally explains the existence of `nanoshots', extremely bright subpulses lasting a few nanoseconds which are sometimes resolved in the Crab's main pulse and low-frequency interpulse \citep[][]{hankins2003, jessner2010, hankins2018}. In this picture, given the conditions in the sheet \citep[]{uzdensky2014}, each nanoshot can be produced by an individual plasmoid merger \citep[][]{lyubarsky2019, philippov2019}, and the microsecond-long `microbursts' that make up the pulses are clumps of many such nanoshots.

Coherent emission from plasmoid mergers is also a candidate emission mechanism for FRBs \citep{lyubarsky2020}. In this scenario, a magnetar flare launches a strong, low-frequency magnetic pulse that propagates outward into the magnetar wind. The pulse compresses the current sheet of the magnetar wind and drives forced reconnection, and the resulting plasmoid mergers produce FMS waves that escape as a coherent radio burst \citep[][]{lyubarsky2020, mahlmann2022}. Coherent radio emission from reconnection has also been predicted as an electromagnetic precursor to compact object mergers, arising from current sheets in the joint magnetospheres of neutron star binaries \citep[][]{most2023, parsons2026} and black hole--neutron star systems \citep[][]{most2023bh} before the merger.

Despite its relevance in a wide variety of astrophysical contexts, numerical studies of this mechanism have, to date, only been performed in 2D \citep[][]{philippov2019, mahlmann2022}. In 3D, the dynamics of relativistic reconnection can differ significantly, as two instabilities arise that are absent in 2D. The first is the relativistic drift-kink instability (RDKI), a primary instability of the initial current sheet that ripples it along the direction of the electric current \citep[][]{zenitani2007,zenitani2008,sironi2014, sironi2025}. The second is the flux-rope kink instability, which causes flux ropes 
to kink once they grow sufficiently large \citep[][]{sironi2014, werner2021, zhang2021, sironi2025}. By limiting the coherence of flux ropes along the current direction, both instabilities may suppress the coherent emission produced by their mergers. Moreover, previous 2D simulations of this mechanism have employed periodic boundary conditions, which provide an environment favorable for plasmoid mergers and may thus overestimate the efficiency of converting incoming Poynting flux into outgoing coherent emission. Finally, previous 2D simulations assumed a vanishing guide field. This is justified for the equatorial current sheets of pulsars, but a guide field may be present in other contexts where this emission mechanism has been invoked, including FRB production and pre-merger binary magnetospheres \citep[]{lyubarsky2020,most2023, most2023bh, parsons2026}. Whether coherent emission from plasmoid mergers remains robust in more realistic setups has therefore remained an open question. A 3D treatment is also necessary to study the polarization of the emitted waves. These waves are linearly polarized, and, in 2D without a guide field, their oscillating electric field is constrained to lie along the current direction by construction. In 3D, flux ropes can kink and merge at oblique angles, so the polarization angle (PA) is no longer constrained by the geometry. This is important observationally, since PA measurements of the Crab’s microbursts and nanoshots, and of FRBs, show a wide range of behaviors, from flat PAs to swings of tens of degrees \citep[][]{jessner2010, hankins2018, panhi2024, ng2025}.

In this work, we study the production of coherent emission from merging flux ropes in 3D relativistic reconnection. We perform large-scale 3D PIC simulations of relativistic reconnection in pair plasma with outflow boundary conditions, for guide-field strengths $B_g/B_0=\{0,0.1,0.3\}$. Despite the richer dynamics in 3D, mergers between flux ropes still occur and launch coherent waves in all guide-field strengths surveyed. We describe our simulation setup in Section \ref{sec:simulation} and our results in Section \ref{sec:results}. In Section \ref{sec:efficiency}, we measure the efficiency with which coherent waves are produced and find that it is remarkably comparable to that obtained in 2D with periodic boundaries. In Section \ref{sec:exhausts}, we propose a physical picture for why the efficiency remains comparable, despite 3D instabilities and our outflow boundary conditions. We then study the PA of the emitted waves in Section \ref{sec:pol}. Finally, in Section \ref{sec:obs}, we summarize our results and discuss observational implications for coherent radio emission from pulsars such as the Crab and for FRBs.

\begin{figure*}
    \centering
    \includegraphics[width=0.99\linewidth]{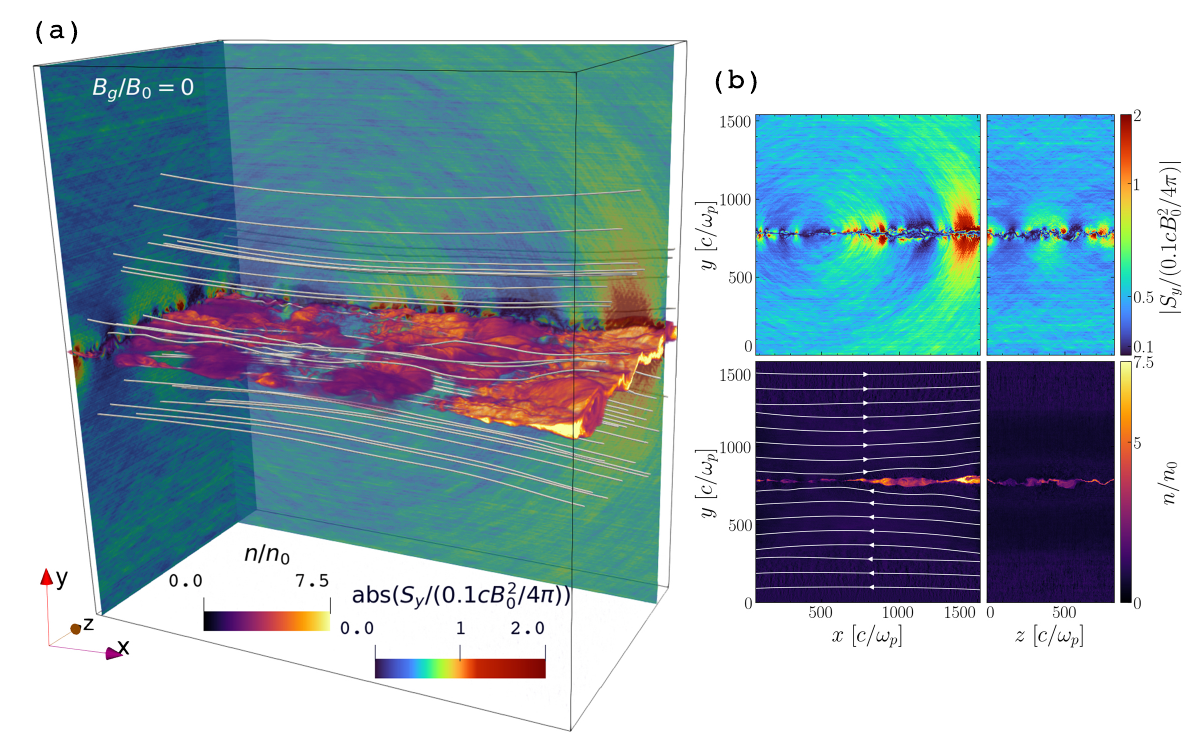}
    \caption{Coherent waves launched by flux rope mergers in the $B_g/B_0 = 0$ simulation, at $ct/L = 6.75$. \textit{Panel (a)}: Volume rendering of particle density, showing flux ropes of various sizes produced by the tearing instability. Two semi-transparent slices, an $x$--$y$ plane near the $z$ boundary and a $z$--$y$ plane near the $x$ boundary, both showing the normalized Poynting flux in $y$, $|S_y/(0.1cB_0^2/4\pi)|$, are also plotted. The faint horizontal striping is an artifact of the moving injector (see Section \ref{sec:efficiency}), and does not affect our efficiency measurement. A selection of magnetic field lines near the sheet is also plotted.
    \textit{Panel (b)}: Normalized Poynting flux in $y$ (top row) and density (bottom row) plotted on the same $x$--$y$ plane (left column) and $z$--$y$ plane (right column) as shown in panel (a). The $x$--$y$ density plot also shows magnetic field lines, which reverse across the sheet. An animation of this figure, for $ct/L\sim5.2\text{--}7.2$, is available at \url{https://youtu.be/WzL2nwgc1Xw}.} 
    \label{fig:waves}
\end{figure*}

\section{Simulation Setup} \label{sec:simulation}

We simulate relativistic reconnection in an electron-positron plasma with the PIC code \texttt{TRISTAN-MP} \citep{buneman1993, spitkovsky2005}. The reconnecting field follows a Harris profile, $B_x = B_0\,\mathrm{tanh}(y/\Delta)$, where $\Delta$ is the half-thickness of the current layer. The guide field points along $z$, and we consider three guide-field strengths, $B_g/B_0 = \{0,\,0.1,\,0.3$\}. The upstream plasma is cold ($k_BT/mc^2 = 10^{-4}$) with number density $n_0$, where $n_0$ is sampled by 2 macroparticles per cell (including both species), and its magnetization is $\sigma=B_0^2/4\pi n_0 mc^2 = (\omega_c/\omega_p)^2 = 10$. Here, $\omega_c = eB_0/mc$ is the gyrofrequency and $\omega_p = \sqrt{4\pi n_0 e^2/m}$ is the plasma frequency. We trigger reconnection by locally reducing the particle pressure at the center of the sheet, launching two fronts towards the $x$ boundaries and clearing out the plasma initialized in the sheet within a crossing time of the box. We define the crossing time of the box as $t = L/c$, where $L=L_x/2=800\,c/\omega_p$. These simulations do not include radiative cooling.

The plasma skin depth $c/\omega_p$ is resolved with $2.5$ cells, and our speed of light is $c=0.45$ cells/timestep. The computational box measures $L_x = 4000\,\mathrm{cells} = 1600\,c/\omega_p$ in the outflow direction and $L_z = 2040\,\mathrm{cells}=816\,c/\omega_p$ in the current (and guide-field) direction. We employ outflow boundary conditions along $x$ and periodic boundary conditions along $z$ \citep[see, e.g.,][]{sironi2016}. Fresh plasma and magnetic flux are introduced into the box by two injectors which recede at the speed of light along $y$. The efficiency measurement in Section \ref{sec:efficiency} is performed on $x$--$z$ planes $\sim500\,c/\omega_p$ from the current sheet, while the polarization angle analysis in Section \ref{sec:pol} is performed on $x$--$z$ planes $\sim640\,c/\omega_p$ from the current sheet. The simulations are run for $\sim7.2$ light-crossing times of the box, and our results in this work are from the time period $ct/L\sim5.2\text{--}7.2$, when the sheet is fully developed. Finally, to avoid any potential contamination by the outflow boundary conditions in $x$, we exclude $80\,c/\omega_p$ from the $x$ edges of the box in all the analyses and figures described in this paper.

\begin{figure*}
    \centering
    \includegraphics[width=0.99\linewidth]{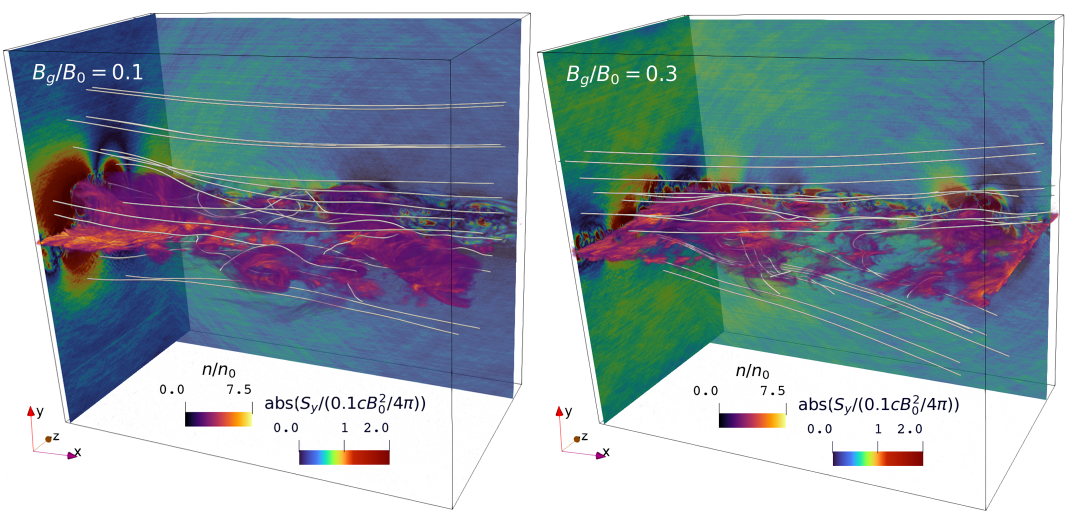}
    \caption{Coherent waves launched by flux rope mergers for $B_g/B_0 = 0.1$ (left) and $B_g/B_0=0.3$ (right), at $ct/L=6.75$. Similarly to Figure \ref{fig:waves}, the volume rendering shows the density, while two semi-transparent slices, an $x$--$y$ plane near the $z$ boundary and a $z$--$y$ plane near the $x$ boundary, show the normalized Poynting flux in the $y$-direction, $|S_y/(0.1cB_0^2/4\pi)|$. A selection of magnetic field lines near the sheet is also plotted in white.}
    \label{fig:waves2}
\end{figure*}

\section{Results} \label{sec:results}

\subsection{Efficiency} \label{sec:efficiency}

In order to study the outgoing waves produced by flux rope mergers in the current sheet, we define the Poynting flux in the $y\text{-}$direction as

\begin{equation}
    S_y = \frac{c}{4\pi}(\mathbf{E}\times\mathbf{B})_y = \frac{c}{4\pi}\big(E_zB_x - E_xB_z\big).
\end{equation}

\noindent This flux includes two distinct contributions: the steady-state reconnection inflow heading towards the sheet, $S_0 \simeq \beta_{\mathrm{rec}}cB_0^2/4\pi$, and the outgoing coherent waves leaving the box.

In Figure \ref{fig:waves}, we show coherent waves launched from flux rope mergers in our $B_g/B_0=0$ simulation. Panel (a) shows a volume rendering of the density at $ct/L=6.75$, with two semi-transparent 2D slices cutting through the box near the $x$ and $z$ boundaries showing $S_y$ normalized by $0.1cB_0^2/4\pi$, i.e., the inflow Poynting flux $S_0$ with a fiducial reconnection rate of $\beta_{\rm{rec}} = 0.1$. The same planes are shown in panel (b): the top row shows $S_y$ again, and the bottom row shows the density for the $x$--$y$ slice (left column) and $z$--$y$ slice (right column). The same volume rendering of the particle density, with semi-transparent $x$--$y$ and $z$--$y$ planes showing the Poynting flux in $y$, is shown in Figure \ref{fig:waves2} for $B_g/B_0 = 0.1$ (left) and $B_g/B_0=0.3$ (right). On these planes, outgoing waves appear as arc-like fronts moving away from the current sheet. These arcs are cross-sections of quasi-spherical wavefronts launched by flux rope mergers. We include an animation of Figure \ref{fig:waves} wherein these waves can be seen propagating out of the box.

To quantify the fraction of the inflowing electromagnetic energy that is carried away by these waves, we measure $S_y$ through two planes oriented perpendicular to the inflow direction, one on each side of the sheet and $\sim500\,c/\omega_p$ away from it. These $x$--$z$ planes intercept the escaping waves without crossing the sheet or its outflows. On each plane we decompose the fields into a time-averaged component and a fluctuating component, i.e., $E_i = \langle E_i\rangle+\delta E_i$ and $B_i = \langle B_i\rangle+\delta B_i$. The time period over which we average is not the full length of the simulation, but rather between $ct/L\sim5.2\text{--}7.2$, i.e., two light-crossing times of the box late in the simulation, once the initial transient has passed and the sheet has fully developed.

The steady inflow is captured in these mean fields: $\langle E_z\rangle$ is the mean motional field of the upstream plasma drifting towards the sheet, and $\langle B_x\rangle$ is the mean reconnecting field component. The mean Poynting flux associated with this inflow is thus \mbox{$S_0 = |(c/4\pi)\langle E_z\rangle \langle B_x \rangle| \simeq \beta_\mathrm{rec}cB_0^2/4\pi$}, and quantifies the rate at which the magnetic energy of the reconnecting field component is delivered to the layer.

\begin{figure}
    \centering
    \includegraphics[width=\columnwidth]{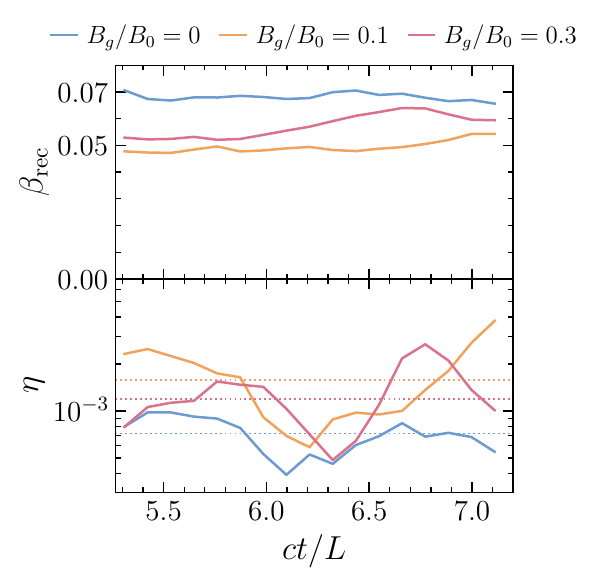}
    \caption{Reconnection rate (top) and efficiency of converting inflowing Poynting flux into outgoing coherent waves (bottom) as a function of
    time. The three simulations, with $B_g/B_0 = \{0, 0.1, 0.3\}$, are shown in blue, orange, and magenta, respectively. Both the rate and the efficiency are measured on $x$--$z$ planes $\sim 500\,c/\omega_p$ from the sheet, and averaged over the two planes. The dotted lines in the bottom panel show the mean efficiency over the whole interval.}
    \label{fig:rate_efficiency}
\end{figure}

The outgoing wave signal is then captured by the fluctuating part of the $y$-component of the Poynting flux, i.e., \mbox{$S_\mathrm{wave} = (c/4\pi)(\delta E_z \delta B_x - \delta E_x \delta B_z)$}. The efficiency of converting inflowing Poynting flux into outgoing coherent waves is thus 

\begin{equation}
    \eta\equiv \frac{S_\mathrm{wave}}{S_0} = \frac{\delta E_z \delta B_x - \delta E_x \delta B_z}{|\langle E_z\rangle\langle B_x\rangle|},
\end{equation}

\noindent where we flip the sign of $S_\mathrm{wave}$ on the lower plane, so that waves leaving the box contribute positively on both planes. We average $\eta$ over each plane at every timestep, then over time, and finally between the two planes. 

For reference, the 2D efficiency measured by \cite{philippov2019} is $\eta\sim1\times10^{-3}$, and, in \cite{mahlmann2022}, $\eta\sim2\times10^{-3}$. Because $S_0$ is measured directly in each simulation, rather than assuming $\beta_\mathrm{rec}\sim0.1$, $\eta$ automatically accounts for variations in the reconnection rate between runs, which allows for direct comparison between different simulations. 

We find the following efficiencies. For our run with zero guide field, $\eta\sim 7\times10^{-4}$, lower by a factor of $\sim2\text{--}3$ than the 2D values. For our moderate guide-field runs we find higher efficiencies than in the 3D zero guide-field case, $\eta\sim2\times10^{-3}$ for $B_g/B_0=0.1$ and $\eta\sim1\times10^{-3}$ for $B_g/B_0=0.3$, similar to previous 2D results.

\begin{figure*}
    \centering
    \includegraphics[width=0.99\linewidth]{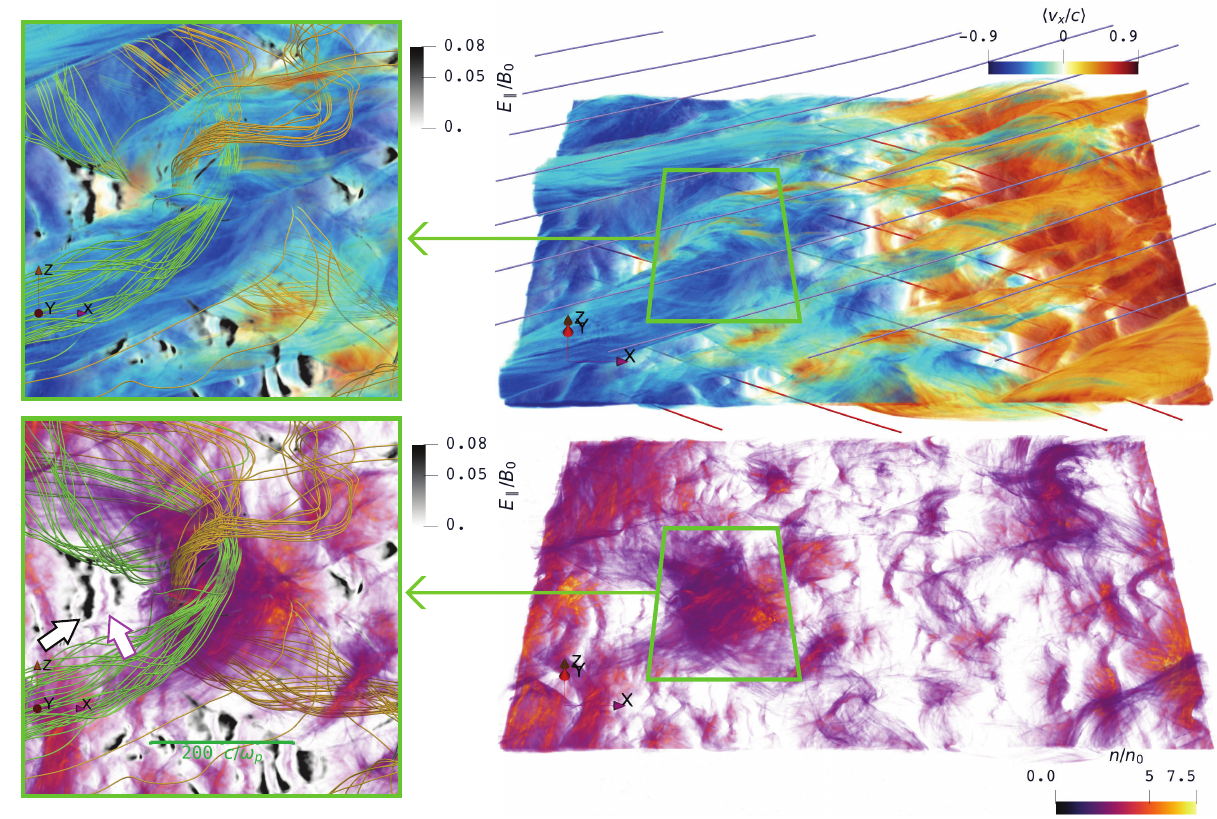}
    \caption{Large-scale, coherent structures in the $B_g/B_0 = 0.3$ simulation at $ct/L = 6.75$. \textit{Top row}: Volume rendering of the average particle outflow velocity $\langle v_x \rangle/c$. Purple and red field lines show the upstream field above and below the sheet, respectively, and the tubes on each side of the sheet follow the corresponding field direction. The left panels zoom in on the interface between a tube above and a tube below the sheet, with a sample of reconnected field lines overlaid. 
    The green and gold field lines originate from two different diffusion regions (shown in black, i.e., regions of strong $E_\parallel)$, and are pulled in opposite directions along $x$. Both of these bundles of field lines cross the sheet, with one end above it and one end below it.
    \textit{Bottom row}: Same as above, but showing particle density. The linkage point between the two bundles appears in the left-hand zoom-in as a large overdense region. The diffusion region that produced the green bundle is indicated by a black-outlined arrow, while a newly created flux rope is indicated by a purple-outlined arrow. A green bar indicates that the density structure is $\sim200\,c/\omega_p$ wide.}
    \label{fig:tubes}
\end{figure*}

These values are robust to the details of the measurement. Shifting the $x$--$z$ planes on which we measure the outgoing wave power, or changing the time interval over which we average, alters $\eta$ by at most a factor of two, with planes farther from the sheet yielding slightly lower values, likely because more of the quasi-spherical waves escape through the open $x$ boundaries before reaching them. We also verified that planar fluctuations from the moving injector, which produce the faint horizontal striping in Figure \ref{fig:waves}, do not contaminate the measurement. Indeed, we ran simulations with different properties of the injector and found the same efficiency. We emphasize that the efficiency is a time-averaged quantity, and because flux rope mergers are by their nature stochastic, time intervals naturally arise during which the efficiency rises by a factor of a few compared to other times with fewer active merger events. Indeed, in Figure \ref{fig:rate_efficiency} we show the instantaneous evolution of the reconnection rate (top) and the efficiency (bottom) for the time interval over which we measure the efficiency, for all three simulations. The reconnection rates for all three guide-field strengths are lower than the fiducial 2D value of $\beta_{\rm{rec}}\sim0.1$, and vary within the range $\beta_{\rm{rec}}\sim0.05\text{--}0.07$, consistent with previous simulations of 3D relativistic reconnection \citep[e.g.,][]{chernoglazov2023, schoeffler2023, hu2026, karavola2026}. Within each run, the reconnection rate remains nearly constant over this time interval, whereas the efficiency varies by a factor of a few, so normalizing by the time-averaged Poynting flux or by its instantaneous value does not affect our results. Note that, at fixed reconnecting field strength, the total power carried away by the waves scales as $\eta\,\beta_{\rm{rec}}$, since $\eta$ is normalized to the inflowing Poynting flux. Thus, although $\eta$ for $B_g/B_0=\{0.1,0.3\}$ matches the 2D value, the wave power is lower than in 2D by up to a factor of two, due to the lower reconnection rate.

The fact that our 3D efficiencies are comparable to those in 2D may at first seem surprising. The RDKI and the flux-rope kink instability, which arise only in 3D, break up the coherent structures whose mergers launch the waves. With outflow boundary conditions, flux ropes can also leave the box without merging at all. A moderate guide field partly counteracts these instabilities, since it stabilizes the sheet against the RDKI \citep[][]{zenitani2008} and raises the minimum unstable wavelength of flux-rope kink modes through the Kruskal–Shafranov criterion \citep[][]{bateman1978, sironi2025}. This, however, cannot explain why the efficiency in all our runs remains comparable to that of 2D simulations with periodic boundaries, despite the different dimensionality and the different choice of boundary conditions. In the next subsection, we examine how reconnected field is transported away from the reconnection sites, and propose a physical picture in which this transport helps sustain the efficiency, particularly in the presence of a moderate guide field.

\subsection{Transport of reconnected field}\label{sec:exhausts}

In the classic 2D reconnection picture without a guide field, a newly reconnected field line lies entirely within an $x$--$y$ plane, and is sharply bent where it crosses the sheet near the X-point. The magnetic tension in this bend then drives the reconnected field away from the X-point, along with its frozen-in plasma. This outflow region, bounded by separatrices on either side of the X-point, forms the reconnection exhaust \citep[see, e.g.,][]{robbins2025}. In the absence of a guide field, this picture largely carries over to 3D, where each field line in the exhaust of a diffusion region remains roughly in the $x$--$y$ plane where it first reconnected. In the presence of even a weak guide field, however, this picture changes. The two legs of the reconnected field lines no longer lie in a common $x$--$y$ plane: they follow the upstream field on either side of the sheet, which is rotated by $\pm\arctan(B_g/B_0)$ from the $x$ axis, and so diverge in $z$. This affects the structure of the outflow \citep[][]{lyubarsky2005}.

In both our 3D simulations with a guide field, we observe that reconnected field lines are organized into distinct, coherent bundles which interact with each other as they flow away from the diffusion region that produced them. These exhausts are best visualized when plotting the average outflow velocity $\langle v_x\rangle/c$ of the particles, which is shown in the top-right panel of Figure \ref{fig:tubes}. There are two distinct contributions to the outflow velocity: the $\bm{E\times B}$ drift, which carries the reconnected field lines away from the diffusion region where they reconnected, and plasma streaming along those field lines. The whole box for the $B_g/B_0 = 0.3$ simulation at $ct/L = 6.75$ is shown, with $\langle v_x\rangle/c$ rendered as a semi-transparent volume so that both sides of the sheet can be seen. In the top-right panel, the purple and red lines show a sample of field lines above and below the sheet, respectively, indicating the geometry of the upstream field. In this render, the exhausts appear as tube-like structures.\footnote{To clarify our terminology, we use ``flux ropes'' for the structures produced by the tearing instability, the 3D analogues of plasmoids, whose field lines wind helically around an axis within the sheet;  ``exhausts'' for the bundles of reconnected field lines flowing away from a diffusion region, which cross the sheet with one end on either side of it; and ``tubes'' for the appearance of exhausts in renderings of $\langle v_x\rangle/c$.} Above the sheet, the tubes follow the purple lines, i.e., the upstream field direction, whereas below the sheet they follow the red lines, whose orientation differs because the reconnecting component reverses across the sheet while the guide field does not. The tubes therefore form a criss-cross pattern. Each tube follows the field because it contains a bundle of reconnected field lines being advected away from the diffusion region that created them, which usually lies near the center of the sheet in $x$. The apex of the bundle, where it bends, sits in the sheet itself, while its two ends lie in tubes on opposite sides of the sheet. In other words, any given tube on one side of the sheet is connected to a tube on the other side, with a tilt between the two that depends on the guide field strength.

Although most of the diffusion regions in our simulations sit near the center of the sheet in $x$, some inflowing field lines reconnect in smaller, off-center diffusion regions, whose exhausts can flow back toward the center. A configuration we observe frequently occurs when such an exhaust becomes topologically linked to an exhaust from the center outflowing in the opposite direction (i.e., towards the boundary of the box), such that the two bundles of reconnected field lines thread through one another like interlocked hooks. The top left zoom-in panel of Figure \ref{fig:tubes} shows one such linked pair. In this zoom-in, we also plot the normalized parallel electric field, $E_\parallel/B_0$. A significant $E_\parallel$ only develops near X-points, where the ideal condition breaks down. There, the reconnecting field vanishes and the overall field strength is dominated by the guide field, which is aligned with the reconnection electric field, so that $E_\parallel \sim \beta_\mathrm{rec}B_0$. Diffusion regions in this plot appear in black. The green field lines were produced by a diffusion region near the center of the sheet and are outflowing in $-\hat{x}$. The gold field lines, on the other hand, were produced by a small diffusion region between the large tubes, which can be seen in black in the middle left of the panel, and are outflowing in $+\hat{x}$. To continue outflowing in $-\hat{x}$, the green bundle must drag the gold bundle along with it, which reduces its outflow velocity. The two bundles begin to reconnect where they meet, but the linked structure as a whole exits the box before the field lines fully reconnect and untangle. The reconnection between the two exhausts is similar to the secondary reconnection that occurs between merging plasmoids in 2D \citep[e.g.,][]{sironi2014}. Unlike plasmoids, however, the exhausts are not necessarily overdense, and the overdensity forms only once they interlock.

The bottom row of Figure \ref{fig:tubes} shows the same snapshot as the top row, except that the rendering illustrates the particle density instead of the average outflow velocity. Regions with densities comparable to or lower than the upstream plasma density are transparent. The exhausts can be either overdense or underdense with respect to the upstream density \citep[]{karavola2026}, which is why they are harder to see in the density render panel. The slowdown of the exhaust of green field lines is what enables the formation of the large density structure at the linkage point between the exhausts, as shown in the bottom left zoom-in panel of Figure \ref{fig:tubes}. Newly created flux ropes (an example of which is indicated by a purple-outlined arrow), mostly generated by the same diffusion region (indicated by a black-outlined arrow) that produced the exhaust of green field lines, catch up to the slow-moving linked structure and merge with it from behind, building up a larger density structure while generating coherent waves. We have confirmed that the largest density structures near the center of the sheet at various times for both $B_g/B_0=0.1$ and $B_g/B_0=0.3$ are typically sites of topological linkage between exhausts which are outflowing in different directions. 

We argue that these local slowdowns of the outflow are central to why the efficiency of coherent wave production in our 3D simulations with outflow boundary conditions is comparable to that of 2D simulations with periodic boundary conditions. In periodic 2D boxes, plasmoids cannot leave the box and are forced to collide head-on, forming large, slow-moving plasmoids into which smaller plasmoids continually merge \citep[]{philippov2019}. In 3D, one might expect the efficiency to drop, because kinked flux ropes can slip past one another rather than merge, and, with outflow boundary conditions, simply exit the box. However, reconnection does not occur only near the center of the sheet: exhausts from off-center diffusion regions interact with the main outflow, creating local overdensities where it slows down. These overdense regions play the role of the large, slow-moving plasmoids in simulations with periodic boundaries. They are also, almost by construction, ideal sites for coherent emission: a slowdown forms where the exhausts of two diffusion regions meet, so it sits between two X-points, which continually produce new flux ropes that their outflows carry directly into the slow-moving structure. Such local slowdowns occur regardless of guide field: we also observe them in our $B_g/B_0 = 0$ run wherever there are off-center diffusion regions, whose exhausts flow against the global outflow, a feature that is not prominent in 2D simulations with outflow boundaries \citep[e.g.,][]{sironi2016}. With a guide field, however, the legs of the reconnected field lines span a $z$-extent that grows with the guide-field strength, so a given diffusion region can slow down the outflow over a wider extent in $z$. Interlocked structures such as the one in Figure \ref{fig:tubes} therefore form more readily, which accounts for the enhanced efficiency we observe at moderate guide fields.

\begin{figure*}
    \centering
    \includegraphics[width=0.99\linewidth]{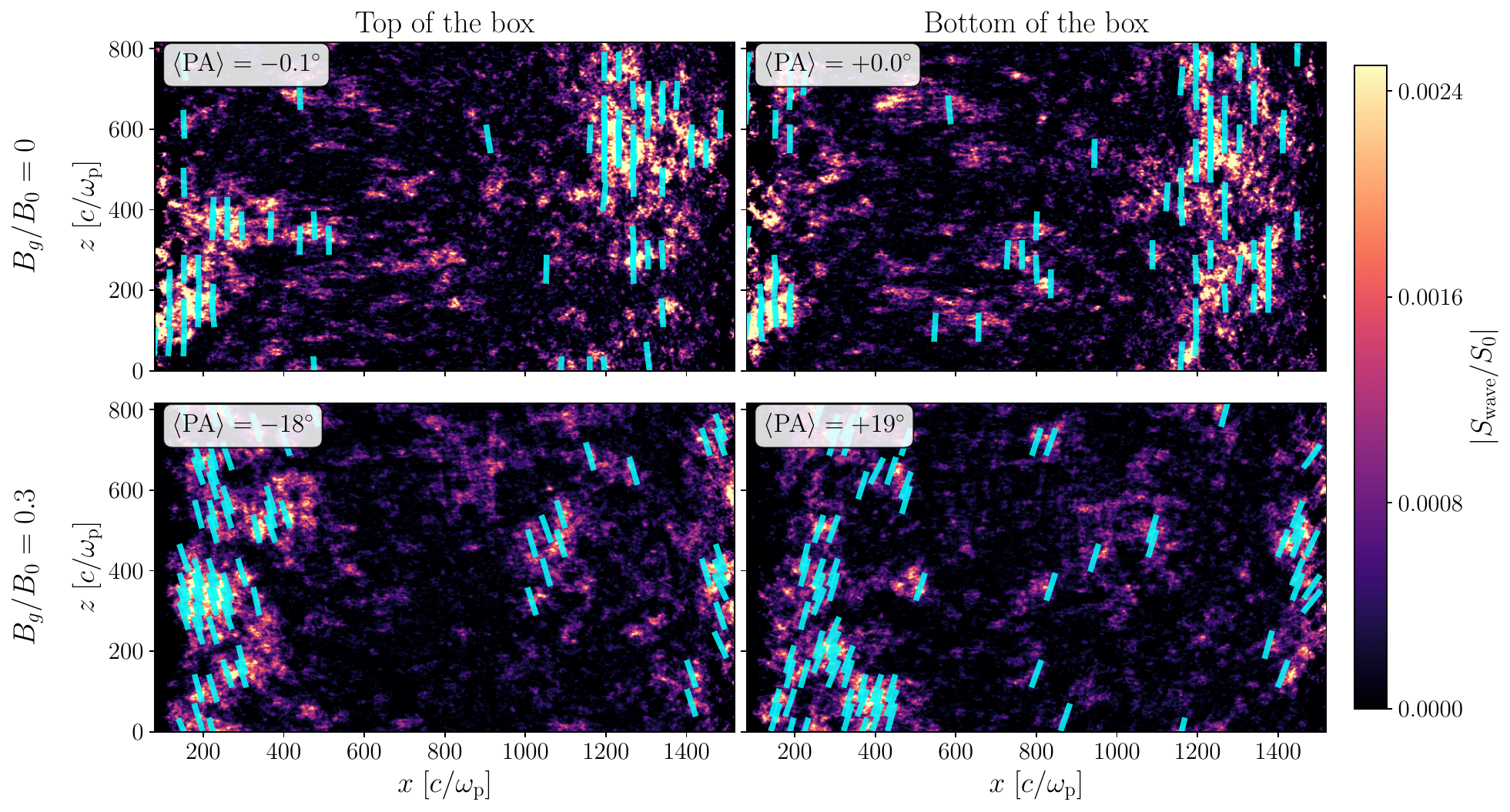}
    \caption{Polarization angle (PA) of the outgoing waves, for $B_g/B_0 = 0$ (top row) and $B_g/B_0 = 0.3$ (bottom row), at $ct/L = 5.6$. The four plots show $x$--$z$ slices near the top of the box (left column) and bottom of the box (right column), $\sim640\,c/\omega_p$ from the current sheet, showing the outgoing Poynting flux carried by the coherent waves, normalized by the inflowing flux $S_0$. The yellow-purple areas are patches of high-intensity waves. The light blue ticks denote the direction of the oscillating electric vector in these bright patches, whose angle from the $z$ axis is the PA. In the $B_g/B_0 = 0$ case, the waves escaping on either side of the box have their electric field aligned with the $z$ axis. For the $B_g/B_0 = 0.3$ case, on the other hand, because the PA of the outgoing waves is set by the direction of the upstream field, there is a rotation across the sheet by $\Delta \mathrm{PA}\sim2\,\arctan(B_g/B_0)$.}
    \label{fig:pol}
\end{figure*}

\subsection{Polarization angle} \label{sec:pol}

We now turn to the polarization angle (PA) of the escaping waves. In all our simulations, we find that these waves are linearly polarized X-mode waves (i.e., with $\delta \bm{E} \perp \bm{B}$), which, as we will show below, sets their polarization angle directly.

At first glance, the fact that the escaping waves all have X-mode polarization may seem surprising. In 2D with vanishing guide field, the oscillating currents driving the waves in a merger event are by necessity along $z$, and so the oscillating electric field of the waves is always perpendicular to the background field. In 3D, however, flux ropes can kink and then merge at oblique angles, and so the oscillating current is no longer constrained to be purely along $z$, which could in principle drive O-mode waves with $\delta \bm{E}\parallel \bm{B}$. However, as we will show below, these O-mode waves cannot escape.

 In a cold magnetized pair plasma, O-mode waves, i.e., waves propagating perpendicular to the background magnetic field with $\delta \bm{E}\parallel \bm{B}$, obey the dispersion relation $\omega^2 = c^2k^2 +\omega_p^2$ \citep[][]{plasma-textbook}, where $\omega_p$ is the plasma frequency. Note that this is the same dispersion relation as in an unmagnetized plasma, because electric field oscillations along $\bm{B}$ are unaffected by the magnetic field. There is thus a cutoff for O-mode waves at $\omega = \omega_p$, i.e., waves below the plasma frequency cannot propagate.

We expect plasmoid mergers to produce waves below this cutoff, i.e., with $\omega < \omega_p$. This is because the smallest flux ropes produced near the X-points have a characteristic size of a few skin depths, i.e., $\ell \sim\zeta c/\omega_p$, where $\zeta$ is a few. The highest frequency of waves emitted in plasmoid mergers is therefore $\omega\sim c/\ell\sim \omega_p/\zeta$, because the size of the smallest merging plasmoid determines the frequency of the waves its merger launches. Indeed, the waves emitted by individual plasmoid mergers in \cite{philippov2019} have durations peaking at $\tau\sim 10/\omega_p$, i.e., at frequencies below $\omega_p$.

X-mode waves, which have $\delta \bm{E} \perp \bm{B}$, behave like fast magnetosonic waves in the highly magnetized limit \citep[see][]{lyubarsky2019, philippov2019}, and so are able to escape the box. In a real magnetosphere, these waves would continue outward through the decreasing plasma density, eventually becoming freely propagating electromagnetic waves. Their PA would become fixed at the polarization-limiting radius, beyond which the plasma is too tenuous for the polarization to follow the local magnetic field \citep[][]{cheng1979, arons1986}. In sum, while flux rope mergers may in principle produce O-mode waves, the only waves to escape the box are X-mode. Note that this differs from the relativistic magnetized shock mechanism for coherent radio emission, where the waves are produced at the shock front, which moves ultra-relativistically relative to the upstream plasma ahead of it \citep[][]{plotnikov2019}. In the rest frame of the upstream plasma, the waves are therefore Doppler-boosted well above the plasma frequency, and both X-mode and O-mode waves can propagate \citep[][]{sironi2021prl,qu2026}.

The fact that, for waves produced by plasmoid mergers, only the X-mode can propagate implies that the polarization angle of the escaping waves (i.e., the orientation of the oscillating electric field) is set by the direction of the upstream magnetic field. Figure \ref{fig:pol} shows this directly. We plot the normalized outgoing Poynting flux of the coherent waves on $x$--$z$ planes near the top (left column) and bottom (right column) of the box. The bright patches are locations of high wave intensity. The top row refers to the $B_g/B_0 = 0$ simulation, and the bottom row is from the $B_g/B_0 = 0.3$ simulation. Light blue ticks show the direction of $\delta \bm{E}$ on the plane and are plotted only for high-intensity patches of $S_\mathrm{wave}$.

For $B_g/B_0 = 0$, the ticks are vertical on both planes, which means that $\delta\bm{E}$ always lies along $z$, perpendicular to the reconnecting field. In the presence of a guide field, however, the upstream field is tilted away from $\hat{x}$ in the $x$--$z$ plane, and because the escaping waves are X-mode, $\delta\bm{E}$ remains perpendicular to the upstream field, and so tilts accordingly. The average PA of the high-intensity patches shown in Figure \ref{fig:pol} is then $\langle\mathrm{PA}\rangle_{\mathrm{bottom}} \approx -\langle\mathrm{PA}\rangle_{\mathrm{top}} \approx \arctan(B_g/B_0)$, where we measure the PA from the $z$ axis. There is thus a flip in PA across the sheet of $\Delta\mathrm{PA} \sim 2\arctan(B_g/B_0)$, set purely by the geometry of the upstream field.

Making detailed observational predictions concerning the PA behavior of the escaping waves requires embedding our simulation in the wind of a pulsar or magnetar, i.e., including geometric and relativistic aberration effects and potential propagation effects as the waves travel through the ambient plasma. If the adiabatic walking condition is satisfied \citep[see, e.g.,][]{cheng1979, arons1986, wang2010}, the polarization will follow the slowly varying background magnetic field as the waves propagate outward, potentially changing the absolute value of the PA while preserving the PA difference imprinted by the guide field under a common large-scale field rotation. If the adiabatic walking condition is not satisfied, the waves may instead retain the polarization angle imprinted at the source \citep[][]{melikidze2014}. While determining the resulting observed PA behavior is the subject of future work, what the present simulations already establish is twofold. First, despite the kinking and turbulent dynamics of the merging flux ropes, the PA of the escaping waves shows very little scatter, at most a few degrees within and between high-intensity patches, and so stochastic PA variability is unlikely to originate from the intrinsic merger dynamics at the source. Second, in the presence of a guide field, waves escaping from opposite sides of the current sheet differ in PA by $\Delta\mathrm{PA} \sim 2\,\arctan(B_g/B_0)$. Both of these findings are relevant for the interpretation of PA variability seen in Crab microbursts and nanoshots, and in FRBs, which we will discuss in Section \ref{sec:obs}.

\section{Summary and Observational Implications} \label{sec:obs}

Despite the more complex dynamics present in 3D relativistic reconnection as compared to 2D, the efficiency with which inflowing Poynting flux is converted into outgoing coherent radiation is largely unchanged. In the absence of a guide field, this efficiency is $\eta\sim7\times10^{-4}$, roughly comparable to the 2D result $\eta\sim10^{-3}$ \citep[]{philippov2019}. The robustness of this efficiency both to 3D dynamics and to outflow (rather than periodic) boundary conditions lends further support to the merging-plasmoid model for coherent radio emission from Crab-like pulsars, where the radio emission is phase-aligned with high-energy emission originating from beyond the light cylinder \citep[][]{lyubarsky2019, philippov2019}.

The efficiency is slightly enhanced at moderate guide fields, yielding $\eta\sim2\times10^{-3}$ for $B_g/B_0=0.1$ and $\eta\sim1\times10^{-3}$ for $B_g/B_0=0.3$. This is important for reconnection-driven models of fast radio bursts \citep[][]{lyubarsky2020, mahlmann2022}, where a magnetar's magnetospheric current sheet is compressed by a low-frequency pulse and the geometry of that compression may not produce a strictly antiparallel field configuration. A nonzero guide field is therefore possible in the FRB setting, and our result that coherent emission remains efficient in that regime supports the model. There has also been recent interest in coherent radio transients produced by relativistic reconnection in the magnetospheres of binary compact objects just before coalescence, such as binary neutron star systems \citep[][]{most2023, parsons2026} and black hole--neutron star binaries \citep[][]{most2023bh}. The current sheets in these scenarios can have moderate guide fields \citep[see][]{parsons2026}, and so the robustness of the efficiency to these guide-field strengths also supports these models. We caution, however, that at strong guide fields the fragmentation of the sheet begins to be significantly affected \citep[][]{zenitani2008, werner2017, hu2026, karavola2026}, and we expect the efficiency to drop primarily due to the slower rate of flux rope formation from tearing. Confirming this requires simulations at higher $B_g/B_0$, which we leave to future work. 

We do not expect our results for the efficiency to be significantly affected by the addition of radiative cooling. Synchrotron cooling reduces the plasma pressure inside the flux ropes, causing them to contract \citep[][]{hakobyan2019, chernoglazov2023}. In 2D, \cite{mahlmann2022} found that cooling reduces plasmoid sizes and shifts the emitted waves to higher frequencies, without affecting the overall dissipation. Since cooled flux ropes are also more strongly magnetized, their mergers may channel a larger fraction of the released energy into coherent waves. This picture remains to be investigated with further 3D simulations of relativistic reconnection, both with and without a guide field, subject to radiative cooling.

We also propose a physical picture for why the efficiency remains comparable to that of 2D simulations with periodic boundaries, despite the 3D instabilities that break up flux ropes and our choice of outflow boundary conditions. Exhausts from off-center diffusion regions interact with the main outflow and slow it down locally. These slow-moving structures play the role of the large plasmoids that form in periodic boxes. Because they sit between diffusion regions, newly formed flux ropes are continually carried into them and merge, launching coherent waves. A guide field enhances this effect. By causing the two legs of each reconnected field line to diverge in $z$, it allows an exhaust to interact with the outflow over a larger portion of the sheet along $z$. Exhausts flowing in opposite directions can then become topologically linked, threading through one another like interlocking hooks, an effect that is intrinsically 3D. This helps explain the higher efficiency at moderate guide fields.

The picture described in Section \ref{sec:exhausts} has consequences beyond the efficiency of coherent emission, which motivate future work. First, reconnection between interlocked bundles provides an additional site of secondary reconnection, analogous to that between merging plasmoids in 2D but occurring between exhausts, and may affect the particle spectrum \citep[see][for a recent study of particle acceleration in 3D relativistic reconnection with a guide field]{karavola2026}. Second, the large, slow-moving structures between linked field bundles may help explain the microbursts observed in the Crab, microsecond-long packets thought to consist of many nanoshots \citep[][]{jessner2010,hankins2018}. In our picture, newly formed flux ropes keep merging into a slow-moving structure for as long as it persists, with each merger producing a nanoshot. The lifetime of such a structure could then set the microsecond duration of a microburst. This picture remains to be confirmed with larger simulations that can follow the lifetime of such structures, which, unlike the duration of a nanoshot, is not set simply by their size. More broadly, our results motivate further study of 3D relativistic reconnection with a guide field, a regime that remains underexplored despite its relevance to black hole magnetospheres and jets \citep[][]{bodo2021, ripperda2022, zhang2022} and magnetar magnetospheres \citep[][]{mahlmann2023}. As we have shown, even a moderate guide field can produce behavior that is absent for $B_g/B_0 = 0$, with direct consequences for the emission from the sheet.

We also find that, because the waves have X-mode polarization, the PA of the waves is set by the direction of the upstream field. In the presence of a guide field, waves escaping from opposite sides of the sheet therefore differ in PA by $\Delta\mathrm{PA}\sim2\,\arctan(B_g/B_0)$. Across single high-intensity patches of emission on either side of the sheet, however, the PA is stable, varying by at most a few degrees, in contrast with the significant PA variability frequently observed in radio emission from the Crab and a subset of FRBs. Indeed, in the Crab, polarization angle measurements across microbursts and nanoshots show highly variable behavior, with some microbursts showing ordered swings of tens of degrees, some showing flat PAs, and some showing seemingly random scatter \citep[][]{jessner2010,hankins2018}. Strikingly, nanoshots can also show swings of tens of degrees over just a few nanoseconds \citep[][]{jessner2010, hankins2018}. The small intrinsic PA variation we see in our simulations is inconsistent with these significant swings. We propose that the rapid variability seen in polarization angle measurements from the Crab's radio emission could instead come from relativistic effects, propagation effects within the pulsar's wind, or passage through structure in the upstream magnetic field, none of which are captured in our simulations. 

Interestingly, PA variability is also seen in FRBs, for both repeaters \citep[][]{luo2020, nimmo2021, ng2025, liu2025} and non-repeaters \citep[][]{panhi2024,mckinven2025, chime2025, balzan2026}. While most FRBs have flat PAs across the burst, others show highly variable and sometimes ordered behavior. The PA swings of a few bursts can be fit with the rotating vector model (RVM), well studied for polar-cap radio emission from pulsars \citep[][]{blaskiewicz1991}. In these cases, however, the inferred rotation periods of the star tend to be shorter than those expected from magnetars \citep[][]{liu2025, mckinven2025}, which are the leading FRB progenitor candidates. This PA behavior is probably intrinsic to the emission, and not caused by the environment, since the same repeater can show significant burst-to-burst variability \citep[]{luo2020, liu2025}. It is worth considering whether physical models other than the RVM, for example ordered flips in PA across a current sheet with a guide field as described in this paper, or other structure in the upstream magnetic field, could explain observed PA swings in FRBs. Indeed, a quantitative model for the observed PA behavior expected from flux rope mergers in a magnetospheric current sheet, including relativistic aberration and propagation effects, is the subject of future work.

\begin{acknowledgments}

The authors would like to thank Matthew Goodbred, Adam Robbins, Vittoria Berta, and Alexander Philippov for interesting discussions which helped improve this manuscript. L.S. acknowledges support from DoE Early Career Award DE-SC0023015, NASA ATP 80NSSC24K1238, NASA ATP 80NSSC24K1826, and NSF AST-2307202. This research was supported by the Multimessenger Plasma Physics Center (MPPC, NSF grants PHY-2206609 and PHY-2206607)  and by a grant from the Simons Foundation (MP-SCMPS-0000147). A.C. acknowledges support from Martin A. and Helen Chooljian Member Fund and the Fund for Memberships in Natural Sciences during his appointment at the IAS.

\end{acknowledgments}

\bibliography{mergingflux}{}

@ARTICLE{lyubarsky2019,
       author = {{Lyubarsky}, Yuri},
        title = "{Radio emission of the Crab and Crab-like pulsars}",
      journal = {\mnras},
         year = 2019,
        month = feb,
       volume = {483},
       number = {2},
        pages = {1731-1736},
          doi = {10.1093/mnras/sty3233},
archivePrefix = {arXiv},
       eprint = {1811.11122},
 primaryClass = {astro-ph.HE},
       adsurl = {https://ui.adsabs.harvard.edu/abs/2019MNRAS.483.1731L}
}

@ARTICLE{philippov2019,
       author = {{Philippov}, Alexander and {Uzdensky}, Dmitri A. and {Spitkovsky}, Anatoly and {Cerutti}, Beno{\^\i}t},
        title = "{Pulsar Radio Emission Mechanism: Radio Nanoshots as a Low-frequency Afterglow of Relativistic Magnetic Reconnection}",
      journal = {\apjl},
         year = 2019,
        month = may,
       volume = {876},
       number = {1},
          eid = {L6},
        pages = {L6},
          doi = {10.3847/2041-8213/ab1590},
archivePrefix = {arXiv},
       eprint = {1902.07730},
 primaryClass = {astro-ph.HE},
       adsurl = {https://ui.adsabs.harvard.edu/abs/2019ApJ...876L...6P}
}

@ARTICLE{uzdensky2014,
       author = {{Uzdensky}, Dmitri A. and {Spitkovsky}, Anatoly},
        title = "{Physical Conditions in the Reconnection Layer in Pulsar Magnetospheres}",
      journal = {\apj},
         year = 2014,
        month = jan,
       volume = {780},
       number = {1},
          eid = {3},
        pages = {3},
          doi = {10.1088/0004-637X/780/1/3},
archivePrefix = {arXiv},
       eprint = {1210.3346},
 primaryClass = {astro-ph.HE},
       adsurl = {https://ui.adsabs.harvard.edu/abs/2014ApJ...780....3U}
}

@ARTICLE{mahlmann2022,
       author = {{Mahlmann}, J.~F. and {Philippov}, A.~A. and {Levinson}, A. and {Spitkovsky}, A. and {Hakobyan}, H.},
        title = "{Electromagnetic Fireworks: Fast Radio Bursts from Rapid Reconnection in the Compressed Magnetar Wind}",
      journal = {\apjl},
         year = 2022,
        month = jun,
       volume = {932},
       number = {2},
          eid = {L20},
        pages = {L20},
          doi = {10.3847/2041-8213/ac7156},
archivePrefix = {arXiv},
       eprint = {2203.04320},
 primaryClass = {astro-ph.HE},
       adsurl = {https://ui.adsabs.harvard.edu/abs/2022ApJ...932L..20M}
}

@ARTICLE{ng2025,
       author = {{Ng}, Cherry and {Pandhi}, Ayush and {Mckinven}, Ryan and {Curtin}, Alice P. and {Shin}, Kaitlyn and {Fonseca}, Emmanuel and {Gaensler}, B.~M. and {Jow}, Dylan L. and {Kaspi}, Victoria and {Li}, Dongzi and {Main}, Robert and {Masui}, Kiyoshi W. and {Michilli}, Daniele and {Nimmo}, Kenzie and {Pleunis}, Ziggy and {Scholz}, Paul and {Stairs}, Ingrid and {Bhardwaj}, Mohit and {Brar}, Charanjot and {Cassanelli}, Tomas and {Joseph}, Ronniy C. and {Pearlman}, Aaron B. and {Rafiei-Ravandi}, Masoud and {Smith}, Kendrick},
        title = "{Polarization Properties of 28 Repeating Fast Radio Burst Sources with CHIME/FRB}",
      journal = {\apj},
         year = 2025,
        month = apr,
       volume = {982},
       number = {2},
          eid = {154},
        pages = {154},
          doi = {10.3847/1538-4357/adb0bc},
archivePrefix = {arXiv},
       eprint = {2411.09045},
 primaryClass = {astro-ph.HE},
       adsurl = {https://ui.adsabs.harvard.edu/abs/2025ApJ...982..154N}
}

@ARTICLE{luo2020,
       author = {{Luo}, R. and {Wang}, B.~J. and {Men}, Y.~P. and {Zhang}, C.~F. and {Jiang}, J.~C. and {Xu}, H. and {Wang}, W.~Y. and {Lee}, K.~J. and {Han}, J.~L. and {Zhang}, B. and {Caballero}, R.~N. and {Chen}, M.~Z. and {Chen}, X.~L. and {Gan}, H.~Q. and {Guo}, Y.~J. and {Hao}, L.~F. and {Huang}, Y.~X. and {Jiang}, P. and {Li}, H. and {Li}, J. and {Li}, Z.~X. and {Luo}, J.~T. and {Pan}, J. and {Pei}, X. and {Qian}, L. and {Sun}, J.~H. and {Wang}, M. and {Wang}, N. and {Wen}, Z.~G. and {Xu}, R.~X. and {Xu}, Y.~H. and {Yan}, J. and {Yan}, W.~M. and {Yu}, D.~J. and {Yuan}, J.~P. and {Zhang}, S.~B. and {Zhu}, Y.},
        title = "{Diverse polarization angle swings from a repeating fast radio burst source}",
      journal = {\nat},
         year = 2020,
        month = oct,
       volume = {586},
       number = {7831},
        pages = {693-696},
          doi = {10.1038/s41586-020-2827-2},
archivePrefix = {arXiv},
       eprint = {2011.00171},
 primaryClass = {astro-ph.HE},
       adsurl = {https://ui.adsabs.harvard.edu/abs/2020Natur.586..693L}
}

@ARTICLE{liu2025,
       author = {{Liu}, Xiaohui and {Xu}, Heng and {Niu}, Jiarui and {Zhang}, Yongkun and {Jiang}, Jinchen and {Zhou}, Dejiang and {Han}, Jinlin and {Zhu}, Weiwei and {Lee}, Kejia and {Li}, Di and {Wang}, Wei-Yang and {Zhang}, Bing and {Chen}, Xuelei and {Luo}, Jia-Wei and {Luo}, Rui and {Niu}, Chenhui and {Qu}, Yuanhong and {Wang}, Bojun and {Wang}, Fayin and {Wang}, Pei and {Wang}, Tiancong and {Wu}, Qin and {Wu}, Ziwei and {Xu}, Jiangwei and {Yang}, Yuan-Pei and {Zhang}, Jun-Shuo},
        title = "{Polarization Position Angle Swing and the Rotating Vector Model of Repeating Fast Radio Bursts}",
      journal = {\apj},
         year = 2025,
        month = aug,
       volume = {988},
       number = {2},
          eid = {175},
        pages = {175},
          doi = {10.3847/1538-4357/ade689},
archivePrefix = {arXiv},
       eprint = {2504.00391},
 primaryClass = {astro-ph.HE},
       adsurl = {https://ui.adsabs.harvard.edu/abs/2025ApJ...988..175L}
}

@ARTICLE{panhi2024,
       author = {{Pandhi}, Ayush and {Pleunis}, Ziggy and {Mckinven}, Ryan and {Gaensler}, B.~M. and {Su}, Jianing and {Ng}, Cherry and {Bhardwaj}, Mohit and {Brar}, Charanjot and {Cassanelli}, Tomas and {Cook}, Amanda and {Curtin}, Alice P. and {Kaspi}, Victoria M. and {Lazda}, Mattias and {Leung}, Calvin and {Li}, Dongzi and {Masui}, Kiyoshi W. and {Michilli}, Daniele and {Nimmo}, Kenzie and {Pearlman}, Aaron B. and {Petroff}, Emily and {Rafiei-Ravandi}, Masoud and {Sand}, Ketan R. and {Scholz}, Paul and {Shin}, Kaitlyn and {Smith}, Kendrick and {Stairs}, Ingrid},
        title = "{Polarization Properties of 128 Nonrepeating Fast Radio Bursts from the First CHIME/FRB Baseband Catalog}",
      journal = {\apj},
         year = 2024,
        month = jun,
       volume = {968},
       number = {2},
          eid = {50},
        pages = {50},
          doi = {10.3847/1538-4357/ad40aa},
archivePrefix = {arXiv},
       eprint = {2401.17378},
 primaryClass = {astro-ph.HE},
       adsurl = {https://ui.adsabs.harvard.edu/abs/2024ApJ...968...50P}
}

@ARTICLE{blaskiewicz1991,
       author = {{Blaskiewicz}, M. and {Cordes}, J.~M. and {Wasserman}, I.},
        title = "{A Relativistic Model of Pulsar Polarization}",
      journal = {\apj},
         year = 1991,
        month = apr,
       volume = {370},
        pages = {643},
          doi = {10.1086/169850},
       adsurl = {https://ui.adsabs.harvard.edu/abs/1991ApJ...370..643B}
}

@ARTICLE{lyubarsky2020,
       author = {{Lyubarsky}, Yuri},
        title = "{Fast Radio Bursts from Reconnection in a Magnetar Magnetosphere}",
      journal = {\apj},
         year = 2020,
        month = jul,
       volume = {897},
       number = {1},
          eid = {1},
        pages = {1},
          doi = {10.3847/1538-4357/ab97b5},
archivePrefix = {arXiv},
       eprint = {2001.02007},
 primaryClass = {astro-ph.HE},
       adsurl = {https://ui.adsabs.harvard.edu/abs/2020ApJ...897....1L}
}

@ARTICLE{chime2025,
       author = {{CHIME/FRB Collaboration} and {Abbott}, Thomas C. and {Amouyal}, Daniel and {Andersen}, Bridget C. and {Andrew}, Shion E. and {Bandura}, Kevin and {Bhardwaj}, Mohit and {Bhopi}, Kalyani and {Bhusare}, Yash and {Brar}, Charanjot and {Cai}, Alice and {Cassanelli}, Tomas and {Chatterjee}, Shami and {Cliche}, Jean-Fran{\c{c}}ois and {Cook}, Amanda M. and {Curtin}, Alice P. and {Davies-Velie}, Evan and {Dobbs}, Matt and {Dong}, Fengqiu Adam and {Dong}, Yuxin and {Eadie}, Gwendolyn and {Eftekhari}, Tarraneh and {Fong}, Wen-Fai and {Fonseca}, Emmanuel and {Gaensler}, B.~M. and {Gusinskaia}, Nina and {Hessels}, Jason W.~T. and {Hewitt}, Dant{\'e} M. and {Huang}, Jeff and {Jain}, Naman and {Joseph}, Ronniy. C. and {Kahinga}, Lordrick and {Kaspi}, Victoria M. and {Khan}, Afrasiyab (Afrokk) and {Kharel}, Bikash and {Lanman}, Adam E. and {L'Argent}, Magnus and {Lazda}, Mattias and {Leung}, Calvin and {Main}, Robert and {Mas-Ribas}, Lluis and {Masui}, Kiyoshi W. and {McGregor}, Kyle and {McKinven}, Ryan and {Mena-Parra}, Juan and {Michilli}, Daniele and {Mulyk}, Nicole and {Ng}, Mason and {Nimmo}, Kenzie and {Pandhi}, Ayush and {Patil}, Swarali Shivraj and {Pearlman}, Aaron B. and {Pen}, Ue-Li and {Pleunis}, Ziggy and {Prochaska}, J. Xavier and {Rafiei-Ravandi}, Masoud and {Ransom}, Scott M. and {Sachdeva}, Gurman and {Sammons}, Mawson W. and {Sand}, Ketan R. and {Scholz}, Paul and {Shah}, Vishwangi and {Shin}, Kaitlyn and {Siegel}, Seth R. and {Simha}, Sunil and {Smith}, Kendrick and {Stairs}, Ingrid and {Stenning}, David C. and {Wang}, Haochen and {Boles}, Thomas and {Cognard}, Isma{\"e}l and {Dijkema}, Tammo Jan and {Filippenko}, Alexei V. and {Gawro{\'n}ski}, Marcin P. and {Herrmann}, Wolfgang and {Kilpatrick}, Charles D. and {Kirsten}, Franz and {Knabel}, Shawn and {Ould-Boukattine}, Omar S. and {Paugnat}, Hadrien and {Puchalska}, Weronika and {Sheu}, William and {Suresh}, Aswin and {Tohuvavohu}, Aaron and {Treu}, Tommaso and {Zheng}, Weikang},
        title = "{FRB 20250316A: A Brilliant and Nearby One-off Fast Radio Burst Localized to 13 pc Precision}",
      journal = {\apjl},
         year = 2025,
        month = aug,
       volume = {989},
       number = {2},
          eid = {L48},
        pages = {L48},
          doi = {10.3847/2041-8213/adf62f},
archivePrefix = {arXiv},
       eprint = {2506.19006},
 primaryClass = {astro-ph.HE},
       adsurl = {https://ui.adsabs.harvard.edu/abs/2025ApJ...989L..48C}
}

@ARTICLE{jessner2010,
       author = {{Jessner}, A. and {Popov}, M.~V. and {Kondratiev}, V.~I. and {Kovalev}, Y.~Y. and {Graham}, D. and {Zensus}, A. and {Soglasnov}, V.~A. and {Bilous}, A.~V. and {Moshkina}, O.~A.},
        title = "{Giant pulses with nanosecond time resolution detected from the Crab pulsar at 8.5 and 15.1 GHz}",
      journal = {\aap},
         year = 2010,
        month = dec,
       volume = {524},
          eid = {A60},
        pages = {A60},
          doi = {10.1051/0004-6361/201014806},
archivePrefix = {arXiv},
       eprint = {1008.3992},
 primaryClass = {astro-ph.HE},
       adsurl = {https://ui.adsabs.harvard.edu/abs/2010A&A...524A..60J}
}

@ARTICLE{hankins2018,
       author = {{Hankins}, T.~H. and {Eilek}, J.~A. and {Jones}, G.},
        title = "{The Crab Pulsar at Centimeter Wavelengths. II. Single Pulses}",
      journal = {\apj},
         year = 2016,
        month = dec,
       volume = {833},
       number = {1},
          eid = {47},
        pages = {47},
          doi = {10.3847/1538-4357/833/1/47},
archivePrefix = {arXiv},
       eprint = {1608.08881},
 primaryClass = {astro-ph.HE},
       adsurl = {https://ui.adsabs.harvard.edu/abs/2016ApJ...833...47H}
}

@ARTICLE{zenitani2008,
       author = {{Zenitani}, S. and {Hoshino}, M.},
        title = "{The Role of the Guide Field in Relativistic Pair Plasma Reconnection}",
      journal = {\apj},
         year = 2008,
        month = apr,
       volume = {677},
       number = {1},
        pages = {530-544},
          doi = {10.1086/528708},
archivePrefix = {arXiv},
       eprint = {0712.2016},
 primaryClass = {astro-ph},
       adsurl = {https://ui.adsabs.harvard.edu/abs/2008ApJ...677..530Z}
}

@ARTICLE{sironi2025,
       author = {{Sironi}, Lorenzo and {Uzdensky}, Dmitri A. and {Giannios}, Dimitrios},
        title = "{Relativistic Magnetic Reconnection in Astrophysical Plasmas: A Powerful Mechanism of Nonthermal Emission}",
      journal = {\araa},
         year = 2025,
        month = aug,
       volume = {63},
       number = {1},
        pages = {127-178},
          doi = {10.1146/annurev-astro-020325-115713},
archivePrefix = {arXiv},
       eprint = {2506.02101},
 primaryClass = {astro-ph.HE},
       adsurl = {https://ui.adsabs.harvard.edu/abs/2025ARA&A..63..127S}
}

@BOOK{plasma-textbook,
       author = {{Swanson}, D. Gary},
        title = "{Plasma Waves, 2nd Edition}",
         year = 2003,
          doi = {10.1201/b15744},
       adsurl = {https://ui.adsabs.harvard.edu/abs/2003plwa.book.....S}
}

@ARTICLE{ripperda2022,
       author = {{Ripperda}, B. and {Liska}, M. and {Chatterjee}, K. and {Musoke}, G. and {Philippov}, A.~A. and {Markoff}, S.~B. and {Tchekhovskoy}, A. and {Younsi}, Z.},
        title = "{Black Hole Flares: Ejection of Accreted Magnetic Flux through 3D Plasmoid-mediated Reconnection}",
      journal = {\apjl},
         year = 2022,
        month = jan,
       volume = {924},
       number = {2},
          eid = {L32},
        pages = {L32},
          doi = {10.3847/2041-8213/ac46a1},
archivePrefix = {arXiv},
       eprint = {2109.15115},
 primaryClass = {astro-ph.HE},
       adsurl = {https://ui.adsabs.harvard.edu/abs/2022ApJ...924L..32R}
}

@ARTICLE{zhang2022,
       author = {{Zhang}, Haocheng and {Li}, Xiaocan and {Giannios}, Dimitrios and {Guo}, Fan and {Thiersen}, Hannes and {B{\"o}ttcher}, Markus and {Lewis}, Tiffany and {Venters}, Tonia},
        title = "{Radiation and Polarization Signatures from Magnetic Reconnection in Relativistic Jets. II. Connection with {\ensuremath{\gamma}}-Rays}",
      journal = {\apj},
         year = 2022,
        month = jan,
       volume = {924},
       number = {2},
          eid = {90},
        pages = {90},
          doi = {10.3847/1538-4357/ac3669},
archivePrefix = {arXiv},
       eprint = {2111.02578},
 primaryClass = {astro-ph.HE},
       adsurl = {https://ui.adsabs.harvard.edu/abs/2022ApJ...924...90Z}
}

@ARTICLE{bodo2021,
       author = {{Bodo}, G. and {Tavecchio}, F. and {Sironi}, L.},
        title = "{Kink-driven magnetic reconnection in relativistic jets: consequences for X-ray polarimetry of BL Lacs}",
      journal = {\mnras},
         year = 2021,
        month = feb,
       volume = {501},
       number = {2},
        pages = {2836-2847},
          doi = {10.1093/mnras/staa3620},
archivePrefix = {arXiv},
       eprint = {2006.14976},
 primaryClass = {astro-ph.HE},
       adsurl = {https://ui.adsabs.harvard.edu/abs/2021MNRAS.501.2836B}
}

@ARTICLE{mahlmann2023,
       author = {{Mahlmann}, J.~F. and {Philippov}, A.~A. and {Mewes}, V. and {Ripperda}, B. and {Most}, E.~R. and {Sironi}, L.},
        title = "{Three-dimensional Dynamics of Strongly Twisted Magnetar Magnetospheres: Kinking Flux Tubes and Global Eruptions}",
      journal = {\apjl},
         year = 2023,
        month = apr,
       volume = {947},
       number = {2},
          eid = {L34},
        pages = {L34},
          doi = {10.3847/2041-8213/accada},
archivePrefix = {arXiv},
       eprint = {2302.07273},
 primaryClass = {astro-ph.HE},
       adsurl = {https://ui.adsabs.harvard.edu/abs/2023ApJ...947L..34M}
}

@ARTICLE{parsons2026,
       author = {{Parsons}, Jasmine and {Spitkovsky}, Anatoly and {Philippov}, Alexander and {Hakobyan}, Hayk},
        title = "{Electromagnetic Precursors to Binary Neutron Star Mergers: Kinetic Simulations of Magnetospheric Flaring}",
      journal = {\apjl},
         year = 2026,
        month = aug,
       volume = {1007},
       number = {2},
          eid = {L50},
        pages = {L50},
          doi = {10.3847/2041-8213/ae9088},
archivePrefix = {arXiv},
       eprint = {2604.22059},
 primaryClass = {astro-ph.HE},
       adsurl = {https://ui.adsabs.harvard.edu/abs/2026ApJ..1007L..50P}
}

@ARTICLE{lyubarsky2005,
       author = {{Lyubarsky}, Y.~E.},
        title = "{On the relativistic magnetic reconnection}",
      journal = {\mnras},
         year = 2005,
        month = mar,
       volume = {358},
       number = {1},
        pages = {113-119},
          doi = {10.1111/j.1365-2966.2005.08767.x},
archivePrefix = {arXiv},
       eprint = {astro-ph/0501392},
 primaryClass = {astro-ph},
       adsurl = {https://ui.adsabs.harvard.edu/abs/2005MNRAS.358..113L}
}

@BOOK{bateman1978,
       author = {{Bateman}, G.},
        title = "{MHD instabilities}",
         year = 1978,
       adsurl = {https://ui.adsabs.harvard.edu/abs/1978mit..book.....B}
}

@ARTICLE{werner2017,
       author = {{Werner}, Gregory R. and {Uzdensky}, Dmitri A.},
        title = "{Nonthermal Particle Acceleration in 3D Relativistic Magnetic Reconnection in Pair Plasma}",
      journal = {\apjl},
         year = 2017,
        month = jul,
       volume = {843},
       number = {2},
          eid = {L27},
        pages = {L27},
          doi = {10.3847/2041-8213/aa7892},
archivePrefix = {arXiv},
       eprint = {1705.05507},
 primaryClass = {astro-ph.HE},
       adsurl = {https://ui.adsabs.harvard.edu/abs/2017ApJ...843L..27W}
}

@ARTICLE{chernoglazov2023,
       author = {{Chernoglazov}, Alexander and {Hakobyan}, Hayk and {Philippov}, Alexander},
        title = "{High-energy Radiation and Ion Acceleration in Three-dimensional Relativistic Magnetic Reconnection with Strong Synchrotron Cooling}",
      journal = {\apj},
         year = 2023,
        month = dec,
       volume = {959},
       number = {2},
          eid = {122},
        pages = {122},
          doi = {10.3847/1538-4357/acffc6},
archivePrefix = {arXiv},
       eprint = {2305.02348},
 primaryClass = {astro-ph.HE},
       adsurl = {https://ui.adsabs.harvard.edu/abs/2023ApJ...959..122C}
}

@ARTICLE{hakobyan2019,
       author = {{Hakobyan}, Hayk and {Philippov}, Alexander and {Spitkovsky}, Anatoly},
        title = "{Effects of Synchrotron Cooling and Pair Production on Collisionless Relativistic Reconnection}",
      journal = {\apj},
         year = 2019,
        month = may,
       volume = {877},
       number = {1},
          eid = {53},
        pages = {53},
          doi = {10.3847/1538-4357/ab191b},
archivePrefix = {arXiv},
       eprint = {1809.10772},
 primaryClass = {astro-ph.HE},
       adsurl = {https://ui.adsabs.harvard.edu/abs/2019ApJ...877...53H}
}

@ARTICLE{nimmo2021,
       author = {{Nimmo}, K. and {Hessels}, J.~W.~T. and {Keimpema}, A. and {Archibald}, A.~M. and {Cordes}, J.~M. and {Karuppusamy}, R. and {Kirsten}, F. and {Li}, D.~Z. and {Marcote}, B. and {Paragi}, Z.},
        title = "{Highly polarized microstructure from the repeating FRB 20180916B}",
      journal = {Nature Astronomy},
         year = 2021,
        month = jun,
       volume = {5},
        pages = {594-603},
          doi = {10.1038/s41550-021-01321-3},
archivePrefix = {arXiv},
       eprint = {2010.05800},
 primaryClass = {astro-ph.HE},
       adsurl = {https://ui.adsabs.harvard.edu/abs/2021NatAs...5..594N}
}

@article{mckinven2025, title={A pulsar-like polarization angle swing from a nearby fast radio burst}, volume={637}, ISSN={0028-0836, 1476-4687}, DOI={10.1038/s41586-024-08184-4}, number={8044}, journal={Nature}, author={Mckinven, Ryan and Bhardwaj, Mohit and Eftekhari, Tarraneh and Kilpatrick, Charles D. and Kirichenko, Aida and Pal, Arpan and Cook, Amanda M. and Gaensler, B. M. and Giri, Utkarsh and Kaspi, Victoria M. and Michilli, Daniele and Nimmo, Kenzie and Pearlman, Aaron B. and Pleunis, Ziggy and Sand, Ketan R. and Stairs, Ingrid and Andersen, Bridget C. and Andrew, Shion and Bandura, Kevin and Brar, Charanjot and Cassanelli, Tomas and Chatterjee, Shami and Curtin, Alice P. and Dong, Fengqiu Adam and Eadie, Gwendolyn and Fonseca, Emmanuel and Ibik, Adaeze L. and Kaczmarek, Jane F. and Kharel, Bikash and Lazda, Mattias and Leung, Calvin and Li, Dongzi and Main, Robert and Masui, Kiyoshi W. and Mena-Parra, Juan and Ng, Cherry and Pandhi, Ayush and Patil, Swarali Shivraj and Prochaska, J. Xavier and Rafiei-Ravandi, Masoud and Scholz, Paul and Shah, Vishwangi and Shin, Kaitlyn and Smith, Kendrick}, year={2025}, month=jan, pages={43–47}, language={en} }

@ARTICLE{most2023,
       author = {{Most}, Elias R. and {Philippov}, Alexander A.},
        title = "{Reconnection-Powered Fast Radio Transients from Coalescing Neutron Star Binaries}",
      journal = {\prl},
         year = 2023,
        month = jun,
       volume = {130},
       number = {24},
          eid = {245201},
        pages = {245201},
          doi = {10.1103/PhysRevLett.130.245201},
archivePrefix = {arXiv},
       eprint = {2207.14435},
 primaryClass = {astro-ph.HE},
       adsurl = {https://ui.adsabs.harvard.edu/abs/2023PhRvL.130x5201M}
}

@ARTICLE{most2023bh,
       author = {{Most}, Elias R. and {Philippov}, Alexander A.},
        title = "{Electromagnetic Precursors to Black Hole-Neutron Star Gravitational Wave Events: Flares and Reconnection-powered Fast Radio Transients from the Late Inspiral}",
      journal = {\apjl},
         year = 2023,
        month = oct,
       volume = {956},
       number = {2},
          eid = {L33},
        pages = {L33},
          doi = {10.3847/2041-8213/acfdae},
archivePrefix = {arXiv},
       eprint = {2309.04271},
 primaryClass = {astro-ph.HE},
       adsurl = {https://ui.adsabs.harvard.edu/abs/2023ApJ...956L..33M}
}

@ARTICLE{loureiro2007,
       author = {{Loureiro}, N.~F. and {Schekochihin}, A.~A. and {Cowley}, S.~C.},
        title = "{Instability of current sheets and formation of plasmoid chains}",
      journal = {Physics of Plasmas},
         year = 2007,
        month = oct,
       volume = {14},
       number = {10},
        pages = {100703-100703},
          doi = {10.1063/1.2783986},
archivePrefix = {arXiv},
       eprint = {astro-ph/0703631},
 primaryClass = {astro-ph},
       adsurl = {https://ui.adsabs.harvard.edu/abs/2007PhPl...14j0703L}
}

@ARTICLE{cerutti2016,
       author = {{Cerutti}, Beno{\^\i}t and {Philippov}, Alexander A. and {Spitkovsky}, Anatoly},
        title = "{Modelling high-energy pulsar light curves from first principles}",
      journal = {\mnras},
         year = 2016,
        month = apr,
       volume = {457},
       number = {3},
        pages = {2401-2414},
          doi = {10.1093/mnras/stw124},
archivePrefix = {arXiv},
       eprint = {1511.01785},
 primaryClass = {astro-ph.HE},
       adsurl = {https://ui.adsabs.harvard.edu/abs/2016MNRAS.457.2401C}
}

@ARTICLE{philippov2018,
       author = {{Philippov}, Alexander A. and {Spitkovsky}, Anatoly},
        title = "{Ab-initio Pulsar Magnetosphere: Particle Acceleration in Oblique Rotators and High-energy Emission Modeling}",
      journal = {\apj},
         year = 2018,
        month = mar,
       volume = {855},
       number = {2},
          eid = {94},
        pages = {94},
          doi = {10.3847/1538-4357/aaabbc},
archivePrefix = {arXiv},
       eprint = {1707.04323},
 primaryClass = {astro-ph.HE},
       adsurl = {https://ui.adsabs.harvard.edu/abs/2018ApJ...855...94P}
}

@ARTICLE{hankins2003,
       author = {{Hankins}, T.~H. and {Kern}, J.~S. and {Weatherall}, J.~C. and {Eilek}, J.~A.},
        title = "{Nanosecond radio bursts from strong plasma turbulence in the Crab pulsar}",
      journal = {\nat},
         year = 2003,
        month = mar,
       volume = {422},
       number = {6928},
        pages = {141-143},
          doi = {10.1038/nature01477},
       adsurl = {https://ui.adsabs.harvard.edu/abs/2003Natur.422..141H}
}

@incollection{buneman1993,
  author    = {Buneman, O.},
  title     = {Simulation Techniques and Software},
  booktitle = {Computer Space Plasma Physics: Simulation Techniques and Software},
  year      = {1993}
}

@INPROCEEDINGS{spitkovsky2005,
       author = {{Spitkovsky}, Anatoly},
        title = "{Simulations of relativistic collisionless shocks: shock structure and particle acceleration}",
    booktitle = {Astrophysical Sources of High Energy Particles and Radiation},
         year = 2005,
       editor = {{Bulik}, Tomasz and {Rudak}, Bronislaw and {Madejski}, Grzegorz},
       series = {American Institute of Physics Conference Series},
       volume = {801},
        month = nov,
    publisher = {AIP},
        pages = {345-350},
          doi = {10.1063/1.2141897},
archivePrefix = {arXiv},
       eprint = {astro-ph/0603211},
 primaryClass = {astro-ph},
       adsurl = {https://ui.adsabs.harvard.edu/abs/2005AIPC..801..345S}
}

@ARTICLE{philippov2022review,
       author = {{Philippov}, A. and {Kramer}, M.},
        title = "{Pulsar Magnetospheres and Their Radiation}",
      journal = {\araa},
         year = 2022,
        month = aug,
       volume = {60},
        pages = {495-558},
          doi = {10.1146/annurev-astro-052920-112338},
       adsurl = {https://ui.adsabs.harvard.edu/abs/2022ARA&A..60..495P}
}

@ARTICLE{philippov2020prl,
       author = {{Philippov}, Alexander and {Timokhin}, Andrey and {Spitkovsky}, Anatoly},
        title = "{Origin of Pulsar Radio Emission}",
      journal = {\prl},
         year = 2020,
        month = jun,
       volume = {124},
       number = {24},
          eid = {245101},
        pages = {245101},
          doi = {10.1103/PhysRevLett.124.245101},
archivePrefix = {arXiv},
       eprint = {2001.02236},
 primaryClass = {astro-ph.HE},
       adsurl = {https://ui.adsabs.harvard.edu/abs/2020PhRvL.124x5101P}
}

@ARTICLE{kaspi2017,
       author = {{Kaspi}, Victoria M. and {Beloborodov}, Andrei M.},
        title = "{Magnetars}",
      journal = {\araa},
         year = 2017,
        month = aug,
       volume = {55},
       number = {1},
        pages = {261-301},
          doi = {10.1146/annurev-astro-081915-023329},
archivePrefix = {arXiv},
       eprint = {1703.00068},
 primaryClass = {astro-ph.HE},
       adsurl = {https://ui.adsabs.harvard.edu/abs/2017ARA&A..55..261K}
}

@ARTICLE{chime2019,
       author = {{CHIME/FRB Collaboration} and {Amiri}, M. and {Bandura}, K. and {Bhardwaj}, M. and {Boubel}, P. and {Boyce}, M.~M. and {Boyle}, P.~J. and {Brar}, C. and {Burhanpurkar}, M. and {Chawla}, P. and {Cliche}, J.~F. and {Cubranic}, D. and {Deng}, M. and {Denman}, N. and {Dobbs}, M. and {Fandino}, M. and {Fonseca}, E. and {Gaensler}, B.~M. and {Gilbert}, A.~J. and {Giri}, U. and {Good}, D.~C. and {Halpern}, M. and {Hanna}, D and {Hill}, A.~S. and {Hinshaw}, G. and {H{\"o}fer}, C. and {Josephy}, A. and {Kaspi}, V.~M. and {Landecker}, T.~L. and {Lang}, D.~A. and {Masui}, K.~W. and {Mckinven}, R. and {Mena-Parra}, J. and {Merryfield}, M. and {Milutinovic}, N. and {Moatti}, C. and {Naidu}, A. and {Newburgh}, L.~B. and {Ng}, C. and {Patel}, C. and {Pen}, U. and {Pinsonneault-Marotte}, T. and {Pleunis}, Z. and {Rafiei-Ravandi}, M. and {Ransom}, S.~M. and {Renard}, A. and {Scholz}, P. and {Shaw}, J.~R. and {Siegel}, S.~R. and {Smith}, K.~M. and {Stairs}, I.~H. and {Tendulkar}, S.~P. and {Tretyakov}, I. and {Vanderlinde}, K. and {Yadav}, P.},
        title = "{Observations of fast radio bursts at frequencies down to 400 megahertz}",
      journal = {\nat},
         year = 2019,
        month = jan,
       volume = {566},
       number = {7743},
        pages = {230-234},
          doi = {10.1038/s41586-018-0867-7},
archivePrefix = {arXiv},
       eprint = {1901.04524},
 primaryClass = {astro-ph.HE},
       adsurl = {https://ui.adsabs.harvard.edu/abs/2019Natur.566..230C}
}

@ARTICLE{petroff2019review,
       author = {{Petroff}, E. and {Hessels}, J.~W.~T. and {Lorimer}, D.~R.},
        title = "{Fast radio bursts}",
      journal = {\aapr},
         year = 2019,
        month = dec,
       volume = {27},
       number = {1},
          eid = {4},
        pages = {4},
          doi = {10.1007/s00159-019-0116-6},
archivePrefix = {arXiv},
       eprint = {1904.07947},
 primaryClass = {astro-ph.HE},
       adsurl = {https://ui.adsabs.harvard.edu/abs/2019A&ARv..27....4P}
}

@ARTICLE{chernoglazov2024,
       author = {{Chernoglazov}, Alexander and {Philippov}, Alexander and {Timokhin}, Andrey},
        title = "{Coherence of Multidimensional Pair Production Discharges in Polar Caps of Pulsars}",
      journal = {\apjl},
         year = 2024,
        month = oct,
       volume = {974},
       number = {2},
          eid = {L32},
        pages = {L32},
          doi = {10.3847/2041-8213/ad7e24},
archivePrefix = {arXiv},
       eprint = {2409.15409},
 primaryClass = {astro-ph.HE},
       adsurl = {https://ui.adsabs.harvard.edu/abs/2024ApJ...974L..32C}
}

@ARTICLE{bransgrove2023,
       author = {{Bransgrove}, Ashley and {Beloborodov}, Andrei M. and {Levin}, Yuri},
        title = "{Radio Emission and Electric Gaps in Pulsar Magnetospheres}",
      journal = {\apjl},
         year = 2023,
        month = nov,
       volume = {958},
       number = {1},
          eid = {L9},
        pages = {L9},
          doi = {10.3847/2041-8213/ad0556},
archivePrefix = {arXiv},
       eprint = {2209.11362},
 primaryClass = {astro-ph.HE},
       adsurl = {https://ui.adsabs.harvard.edu/abs/2023ApJ...958L...9B}
}

@ARTICLE{cruz2021,
       author = {{Cruz}, F{\'a}bio and {Grismayer}, Thomas and {Chen}, Alexander Y. and {Spitkovsky}, Anatoly and {Silva}, Luis O.},
        title = "{Coherent Emission from QED Cascades in Pulsar Polar Caps}",
      journal = {\apjl},
         year = 2021,
        month = sep,
       volume = {919},
       number = {1},
          eid = {L4},
        pages = {L4},
          doi = {10.3847/2041-8213/ac2157},
archivePrefix = {arXiv},
       eprint = {2108.11702},
 primaryClass = {astro-ph.HE},
       adsurl = {https://ui.adsabs.harvard.edu/abs/2021ApJ...919L...4C}
}

@ARTICLE{moffett1996,
       author = {{Moffett}, David A. and {Hankins}, Timothy H.},
        title = "{Multifrequency Radio Observations of the Crab Pulsar}",
      journal = {\apj},
         year = 1996,
        month = sep,
       volume = {468},
        pages = {779},
          doi = {10.1086/177734},
archivePrefix = {arXiv},
       eprint = {astro-ph/9604163},
 primaryClass = {astro-ph},
       adsurl = {https://ui.adsabs.harvard.edu/abs/1996ApJ...468..779M}
}

@ARTICLE{johnson2014,
       author = {{Johnson}, T.~J. and {Venter}, C. and {Harding}, A.~K. and {Guillemot}, L. and {Smith}, D.~A. and {Kramer}, M. and {{\c{C}}elik}, {\"O}. and {den Hartog}, P.~R. and {Ferrara}, E.~C. and {Hou}, X. and {Lande}, J. and {Ray}, P.~S.},
        title = "{Constraints on the Emission Geometries and Spin Evolution of Gamma-Ray Millisecond Pulsars}",
      journal = {\apjs},
         year = 2014,
        month = jul,
       volume = {213},
       number = {1},
          eid = {6},
        pages = {6},
          doi = {10.1088/0067-0049/213/1/6},
archivePrefix = {arXiv},
       eprint = {1404.2264},
 primaryClass = {astro-ph.HE},
       adsurl = {https://ui.adsabs.harvard.edu/abs/2014ApJS..213....6J}
}

@ARTICLE{abdo2010,
       author = {{Abdo}, A.~A. and {Ackermann}, M. and {Ajello}, M. and {Atwood}, W.~B. and {Axelsson}, M. and {Baldini}, L. and {Ballet}, J. and {Barbiellini}, G. and {Baring}, M.~G. and {Bastieri}, D. and {Bechtol}, K. and {Bellazzini}, R. and {Berenji}, B. and {Blandford}, R.~D. and {Bloom}, E.~D. and {Bonamente}, E. and {Borgland}, A.~W. and {Bregeon}, J. and {Brez}, A. and {Brigida}, M. and {Bruel}, P. and {Burnett}, T.~H. and {Caliandro}, G.~A. and {Cameron}, R.~A. and {Camilo}, F. and {Caraveo}, P.~A. and {Casandjian}, J.~M. and {Cecchi}, C. and {{\c{C}}elik}, {\"O}. and {Chekhtman}, A. and {Cheung}, C.~C. and {Chiang}, J. and {Ciprini}, S. and {Claus}, R. and {Cognard}, I. and {Cohen-Tanugi}, J. and {Cominsky}, L.~R. and {Conrad}, J. and {Dermer}, C.~D. and {de Angelis}, A. and {de Luca}, A. and {de Palma}, F. and {Digel}, S.~W. and {Silva}, E. do Couto e. and {Drell}, P.~S. and {Dubois}, R. and {Dumora}, D. and {Espinoza}, C. and {Farnier}, C. and {Favuzzi}, C. and {Fegan}, S.~J. and {Ferrara}, E.~C. and {Focke}, W.~B. and {Frailis}, M. and {Freire}, P.~C.~C. and {Fukazawa}, Y. and {Funk}, S. and {Fusco}, P. and {Gargano}, F. and {Gasparrini}, D. and {Gehrels}, N. and {Germani}, S. and {Giavitto}, G. and {Giebels}, B. and {Giglietto}, N. and {Giordano}, F. and {Glanzman}, T. and {Godfrey}, G. and {Grenier}, I.~A. and {Grondin}, M.-H. and {Grove}, J.~E. and {Guillemot}, L. and {Guiriec}, S. and {Hanabata}, Y. and {Harding}, A.~K. and {Hayashida}, M. and {Hays}, E. and {Hughes}, R.~E. and {J{\'o}hannesson}, G. and {Johnson}, A.~S. and {Johnson}, R.~P. and {Johnson}, T.~J. and {Johnson}, W.~N. and {Johnston}, S. and {Kamae}, T. and {Katagiri}, H. and {Kataoka}, J. and {Kawai}, N. and {Kerr}, M. and {Kn{\"o}dlseder}, J. and {Kocian}, M.~L. and {Kramer}, M. and {Kuehn}, F. and {Kuss}, M. and {Lande}, J. and {Latronico}, L. and {Lee}, S.-H. and {Lemoine-Goumard}, M. and {Longo}, F. and {Loparco}, F. and {Lott}, B. and {Lovellette}, M.~N. and {Lubrano}, P. and {Lyne}, A.~G. and {Makeev}, A. and {Marelli}, M. and {Mazziotta}, M.~N. and {McEnery}, J.~E. and {Meurer}, C. and {Michelson}, P.~F. and {Mitthumsiri}, W. and {Mizuno}, T. and {Moiseev}, A.~A. and {Monte}, C. and {Monzani}, M.~E. and {Moretti}, E. and {Morselli}, A. and {Moskalenko}, I.~V. and {Murgia}, S. and {Nakamori}, T. and {Nolan}, P.~L. and {Norris}, J.~P. and {Noutsos}, A. and {Nuss}, E. and {Ohsugi}, T. and {Omodei}, N. and {Orlando}, E. and {Ormes}, J.~F. and {Ozaki}, M. and {Paneque}, D. and {Panetta}, J.~H. and {Parent}, D. and {Pelassa}, V. and {Pepe}, M. and {Pesce-Rollins}, M. and {Pierbattista}, M. and {Piron}, F. and {Porter}, T.~A. and {Rain{\`o}}, S. and {Rando}, R. and {Ray}, P.~S. and {Razzano}, M. and {Reimer}, A. and {Reimer}, O. and {Reposeur}, T. and {Ritz}, S. and {Rochester}, L.~S. and {Rodriguez}, A.~Y. and {Romani}, R.~W. and {Roth}, M. and {Ryde}, F. and {Sadrozinski}, H.~F.-W. and {Sanchez}, D. and {Sander}, A. and {Saz Parkinson}, P.~M. and {Scargle}, J.~D. and {Sgr{\`o}}, C. and {Siskind}, E.~J. and {Smith}, D.~A. and {Smith}, P.~D. and {Spandre}, G. and {Spinelli}, P. and {Stappers}, B.~W. and {Strickman}, M.~S. and {Suson}, D.~J. and {Tajima}, H. and {Takahashi}, H. and {Tanaka}, T. and {Thayer}, J.~B. and {Thayer}, J.~G. and {Theureau}, G. and {Thompson}, D.~J. and {Thorsett}, S.~E. and {Tibaldo}, L. and {Torres}, D.~F. and {Tosti}, G. and {Tramacere}, A. and {Uchiyama}, Y. and {Usher}, T.~L. and {Van Etten}, A. and {Vasileiou}, V. and {Vilchez}, N. and {Vitale}, V. and {Waite}, A.~P. and {Wallace}, E. and {Wang}, P. and {Watters}, K. and {Weltevrede}, P. and {Winer}, B.~L. and {Wood}, K.~S. and {Ylinen}, T. and {Ziegler}, M.},
        title = "{Fermi Large Area Telescope Observations of the Crab Pulsar And Nebula}",
      journal = {\apj},
         year = 2010,
        month = jan,
       volume = {708},
       number = {2},
        pages = {1254-1267},
          doi = {10.1088/0004-637X/708/2/1254},
archivePrefix = {arXiv},
       eprint = {0911.2412},
 primaryClass = {astro-ph.HE},
       adsurl = {https://ui.adsabs.harvard.edu/abs/2010ApJ...708.1254A}
}

@ARTICLE{eilek2016,
       author = {{Eilek}, Jean A. and {Hankins}, Timothy H.},
        title = "{Radio emission physics in the Crab pulsar}",
      journal = {Journal of Plasma Physics},
         year = 2016,
        month = jun,
       volume = {82},
       number = {3},
          eid = {635820302},
        pages = {635820302},
          doi = {10.1017/S002237781600043X},
archivePrefix = {arXiv},
       eprint = {1604.02472},
 primaryClass = {astro-ph.HE},
       adsurl = {https://ui.adsabs.harvard.edu/abs/2016JPlPh..82c6302E}
}

@ARTICLE{guillemot2012,
       author = {{Guillemot}, L. and {Johnson}, T.~J. and {Venter}, C. and {Kerr}, M. and {Pancrazi}, B. and {Livingstone}, M. and {Janssen}, G.~H. and {Jaroenjittichai}, P. and {Kramer}, M. and {Cognard}, I. and {Stappers}, B.~W. and {Harding}, A.~K. and {Camilo}, F. and {Espinoza}, C.~M. and {Freire}, P.~C.~C. and {Gargano}, F. and {Grove}, J.~E. and {Johnston}, S. and {Michelson}, P.~F. and {Noutsos}, A. and {Parent}, D. and {Ransom}, S.~M. and {Ray}, P.~S. and {Shannon}, R. and {Smith}, D.~A. and {Theureau}, G. and {Thorsett}, S.~E. and {Webb}, N.},
        title = "{Pulsed Gamma Rays from the Original Millisecond and Black Widow Pulsars: A Case for Caustic Radio Emission?}",
      journal = {\apj},
         year = 2012,
        month = jan,
       volume = {744},
       number = {1},
          eid = {33},
        pages = {33},
          doi = {10.1088/0004-637X/744/1/33},
archivePrefix = {arXiv},
       eprint = {1110.1271},
 primaryClass = {astro-ph.HE},
       adsurl = {https://ui.adsabs.harvard.edu/abs/2012ApJ...744...33G}
}

@ARTICLE{zenitani2007,
       author = {{Zenitani}, S. and {Hoshino}, M.},
        title = "{Particle Acceleration and Magnetic Dissipation in Relativistic Current Sheet of Pair Plasmas}",
      journal = {\apj},
         year = 2007,
        month = nov,
       volume = {670},
       number = {1},
        pages = {702-726},
          doi = {10.1086/522226},
archivePrefix = {arXiv},
       eprint = {0708.1000},
 primaryClass = {astro-ph},
       adsurl = {https://ui.adsabs.harvard.edu/abs/2007ApJ...670..702Z}
}

@ARTICLE{sironi2014,
       author = {{Sironi}, Lorenzo and {Spitkovsky}, Anatoly},
        title = "{Relativistic Reconnection: An Efficient Source of Non-thermal Particles}",
      journal = {\apjl},
         year = 2014,
        month = mar,
       volume = {783},
       number = {1},
          eid = {L21},
        pages = {L21},
          doi = {10.1088/2041-8205/783/1/L21},
archivePrefix = {arXiv},
       eprint = {1401.5471},
 primaryClass = {astro-ph.HE},
       adsurl = {https://ui.adsabs.harvard.edu/abs/2014ApJ...783L..21S}
}

@ARTICLE{werner2021,
       author = {{Werner}, Gregory R. and {Uzdensky}, Dmitri A.},
        title = "{Reconnection and particle acceleration in three-dimensional current sheet evolution in moderately magnetized astrophysical pair plasma}",
      journal = {Journal of Plasma Physics},
         year = 2021,
        month = dec,
       volume = {87},
       number = {6},
          eid = {905870613},
        pages = {905870613},
          doi = {10.1017/S0022377821001185},
archivePrefix = {arXiv},
       eprint = {2106.02790},
 primaryClass = {astro-ph.HE},
       adsurl = {https://ui.adsabs.harvard.edu/abs/2021JPlPh..87f9013W}
}

@ARTICLE{zhang2021,
       author = {{Zhang}, Hao and {Sironi}, Lorenzo and {Giannios}, Dimitrios},
        title = "{Fast Particle Acceleration in Three-dimensional Relativistic Reconnection}",
      journal = {\apj},
         year = 2021,
        month = dec,
       volume = {922},
       number = {2},
          eid = {261},
        pages = {261},
          doi = {10.3847/1538-4357/ac2e08},
archivePrefix = {arXiv},
       eprint = {2105.00009},
 primaryClass = {astro-ph.HE},
       adsurl = {https://ui.adsabs.harvard.edu/abs/2021ApJ...922..261Z}
}

@ARTICLE{schoeffler2023,
       author = {{Schoeffler}, K.~M. and {Grismayer}, T. and {Uzdensky}, D. and {Silva}, L.~O.},
        title = "{High-energy synchrotron flares powered by strongly radiative relativistic magnetic reconnection: 2D and 3D PIC simulations}",
      journal = {\mnras},
         year = 2023,
        month = aug,
       volume = {523},
       number = {3},
        pages = {3812-3839},
          doi = {10.1093/mnras/stad1588},
archivePrefix = {arXiv},
       eprint = {2303.16643},
 primaryClass = {astro-ph.HE},
       adsurl = {https://ui.adsabs.harvard.edu/abs/2023MNRAS.523.3812S}
}

@ARTICLE{cheng1979,
       author = {{Cheng}, A.~F. and {Ruderman}, M.~A.},
        title = "{A theory of subpulse polarization patterns from radio pulsars.}",
      journal = {\apj},
         year = 1979,
        month = apr,
       volume = {229},
        pages = {348-360},
          doi = {10.1086/156959},
       adsurl = {https://ui.adsabs.harvard.edu/abs/1979ApJ...229..348C}
}

@ARTICLE{wang2010,
       author = {{Wang}, Chen and {Lai}, Dong and {Han}, Jinlin},
        title = "{Polarization changes of pulsars due to wave propagation through magnetospheres}",
      journal = {\mnras},
         year = 2010,
        month = apr,
       volume = {403},
       number = {2},
        pages = {569-588},
          doi = {10.1111/j.1365-2966.2009.16074.x},
archivePrefix = {arXiv},
       eprint = {0910.2793},
 primaryClass = {astro-ph.HE},
       adsurl = {https://ui.adsabs.harvard.edu/abs/2010MNRAS.403..569W}
}

@ARTICLE{arons1986,
       author = {{Arons}, J. and {Barnard}, J.~J.},
        title = "{Wave Propagation in Pulsar Magnetospheres: Dispersion Relations and Normal Modes of Plasmas in Superstrong Magnetic Fields}",
      journal = {\apj},
         year = 1986,
        month = mar,
       volume = {302},
        pages = {120},
          doi = {10.1086/163978},
       adsurl = {https://ui.adsabs.harvard.edu/abs/1986ApJ...302..120A}
}

@ARTICLE{melikidze2014,
       author = {{Melikidze}, George I. and {Mitra}, Dipanjan and {Gil}, Janusz},
        title = "{On the Adiabatic Walking of Plasma Waves in a Pulsar Magnetosphere}",
      journal = {\apj},
         year = 2014,
        month = oct,
       volume = {794},
       number = {2},
          eid = {105},
        pages = {105},
          doi = {10.1088/0004-637X/794/2/105},
archivePrefix = {arXiv},
       eprint = {1409.1587},
 primaryClass = {astro-ph.SR},
       adsurl = {https://ui.adsabs.harvard.edu/abs/2014ApJ...794..105M}
}

@ARTICLE{plotnikov2019,
       author = {{Plotnikov}, Illya and {Sironi}, Lorenzo},
        title = "{The synchrotron maser emission from relativistic shocks in Fast Radio Bursts: 1D PIC simulations of cold pair plasmas}",
      journal = {\mnras},
         year = 2019,
        month = may,
       volume = {485},
       number = {3},
        pages = {3816-3833},
          doi = {10.1093/mnras/stz640},
archivePrefix = {arXiv},
       eprint = {1901.01029},
 primaryClass = {astro-ph.HE},
       adsurl = {https://ui.adsabs.harvard.edu/abs/2019MNRAS.485.3816P}
}

@ARTICLE{qu2026,
       author = {{Qu}, Yuanhong and {N{\"a}ttil{\"a}}, Joonas and {Sironi}, Lorenzo},
        title = "{Spectral and Polarization Properties of Coherent Radio Emission from Relativistic Magnetized Pair Plasma Shocks}",
      journal = {arXiv e-prints},
         year = 2026,
        month = sep,
          eid = {arXiv:2609.30494},
        pages = {arXiv:2609.30494},
          doi = {10.48550/arXiv.2609.30494},
archivePrefix = {arXiv},
       eprint = {2609.30494},
 primaryClass = {astro-ph.HE},
       adsurl = {https://ui.adsabs.harvard.edu/abs/2026arXiv260930494Q}
}

@ARTICLE{karavola2026,
       author = {{Karavola}, Despina and {Sironi}, Lorenzo and {Petropoulou}, Maria},
        title = "{Guide field effects on particle acceleration in three-dimensional relativistic magnetic reconnection}",
      journal = {arXiv e-prints},
         year = 2026,
        month = oct,
          eid = {arXiv:2610.03427},
        pages = {arXiv:2610.03427},
          doi = {10.48550/arXiv.2610.03427},
archivePrefix = {arXiv},
       eprint = {2610.03427},
 primaryClass = {astro-ph.HE},
       adsurl = {https://ui.adsabs.harvard.edu/abs/2026arXiv261003427K}
}

@ARTICLE{robbins2025,
       author = {{Robbins}, Adam and {Spitkovsky}, Anatoly},
        title = "{Transition to Petschek Reconnection in Subrelativistic Pair Plasmas: Implications for Particle Acceleration}",
      journal = {\apj},
         year = 2025,
        month = dec,
       volume = {995},
       number = {2},
          eid = {163},
        pages = {163},
          doi = {10.3847/1538-4357/ae1618},
archivePrefix = {arXiv},
       eprint = {2508.08533},
 primaryClass = {physics.plasm-ph},
       adsurl = {https://ui.adsabs.harvard.edu/abs/2025ApJ...995..163R}
}

@ARTICLE{lorimer2007,
       author = {{Lorimer}, D.~R. and {Bailes}, M. and {McLaughlin}, M.~A. and {Narkevic}, D.~J. and {Crawford}, F.},
        title = "{A Bright Millisecond Radio Burst of Extragalactic Origin}",
      journal = {Science},
         year = 2007,
        month = nov,
       volume = {318},
       number = {5851},
        pages = {777},
          doi = {10.1126/science.1147532},
archivePrefix = {arXiv},
       eprint = {0709.4301},
 primaryClass = {astro-ph},
       adsurl = {https://ui.adsabs.harvard.edu/abs/2007Sci...318..777L}
}

@ARTICLE{sironi2021prl,
       author = {{Sironi}, Lorenzo and {Plotnikov}, Illya and {N{\"a}ttil{\"a}}, Joonas and {Beloborodov}, Andrei M.},
        title = "{Coherent Electromagnetic Emission from Relativistic Magnetized Shocks}",
      journal = {\prl},
         year = 2021,
        month = jul,
       volume = {127},
       number = {3},
          eid = {035101},
        pages = {035101},
          doi = {10.1103/PhysRevLett.127.035101},
archivePrefix = {arXiv},
       eprint = {2107.01211},
 primaryClass = {astro-ph.HE},
       adsurl = {https://ui.adsabs.harvard.edu/abs/2021PhRvL.127c5101S}
}

@ARTICLE{hu2026,
       author = {{Hu}, Yue and {Comisso}, Luca and {Sironi}, Lorenzo and {Xu}, Siyao},
        title = "{Reconnection-driven Turbulent Fluctuations in the Magnetically Dominated Collisionless Regime}",
      journal = {\apj},
         year = 2026,
        month = jul,
       volume = {1005},
       number = {2},
          eid = {197},
        pages = {197},
          doi = {10.3847/1538-4357/ae75b4},
archivePrefix = {arXiv},
       eprint = {2512.12516},
 primaryClass = {astro-ph.GA},
       adsurl = {https://ui.adsabs.harvard.edu/abs/2026ApJ..1005..197H}
}

@ARTICLE{kramer2026,
       author = {{Kramer}, Michael and {Johnston}, Simon},
        title = "{Radio emission from beyond the light cylinder in millisecond pulsars}",
      journal = {\mnras},
         year = 2026,
        month = apr,
       volume = {547},
       number = {4},
        pages = {1-22},
          doi = {10.1093/mnras/staf2258},
archivePrefix = {arXiv},
       eprint = {2510.05778},
 primaryClass = {astro-ph.HE},
       adsurl = {https://ui.adsabs.harvard.edu/abs/2026MNRAS.547f2258K}
}

@ARTICLE{sironi2016,
       author = {{Sironi}, Lorenzo and {Giannios}, Dimitrios and {Petropoulou}, Maria},
        title = "{Plasmoids in relativistic reconnection, from birth to adulthood: first they grow, then they go}",
      journal = {\mnras},
         year = 2016,
        month = oct,
       volume = {462},
       number = {1},
        pages = {48-74},
          doi = {10.1093/mnras/stw1620},
archivePrefix = {arXiv},
       eprint = {1605.02071},
 primaryClass = {astro-ph.HE},
       adsurl = {https://ui.adsabs.harvard.edu/abs/2016MNRAS.462...48S}
}

@ARTICLE{balzan2026,
       author = {{Balzan}, J.~C.~F. and {Bera}, A. and {James}, C.~W. and {Meyers}, B.~W. and {Deller}, A.~T. and {Shannon}, R.~M. and {Wang}, Z.},
        title = "{FRB 20250607A: Weirder Than Fiction}",
      journal = {arXiv e-prints},
         year = 2026,
        month = sep,
          eid = {arXiv:2609.18086},
        pages = {arXiv:2609.18086},
          doi = {10.48550/arXiv.2609.18086},
archivePrefix = {arXiv},
       eprint = {2609.18086},
 primaryClass = {astro-ph.HE},
       adsurl = {https://ui.adsabs.harvard.edu/abs/2026arXiv260918086B}
}
\bibliographystyle{aasjournalv7}

\end{document}